\documentclass[twocolumn,graphics,floatfix,a4paper,prx]{revtex4-1}
\usepackage{color}
\usepackage[utf8]{inputenc}
\usepackage{graphicx}
\usepackage{amsmath}
\usepackage{amssymb}
\usepackage{longtable}

\usepackage{hyperref}
\hypersetup{
  pdfnewwindow=true, colorlinks=true,
  linkcolor=blue, anchorcolor=blue,
  citecolor=blue, filecolor=blue,
  menucolor=blue, urlcolor=blue}

\graphicspath{{./Fig/}}

\newcommand{\comment}[1]{\textit{}}
\newcommand{\bit}{\begin{itemize} \setlength{\itemsep}{0ex} \setlength{\topsep}{0ex} } 
\newcommand{\eit}{\end{itemize}}
\newcommand{\be}{\begin{equation}}
\newcommand{\ee}{\end{equation}}
\newcommand{\bea}{\begin{eqnarray}}
\newcommand{\eea}{\end{eqnarray}}
\newcommand{\ba}{\begin{align}}
\newcommand{\ea}{\end{align}}
\newcommand{\SKIP}[1]{}

\providecommand{\bk}{\ensuremath{{\boldsymbol k}}}
\providecommand{\kp}{\ensuremath{{\boldsymbol k}\cdot{\boldsymbol p}}}

\begin{document}
\author{Ziming Zhu$^{1,3}$}
\author{Georg W. Winkler$^{2}$}
\thanks{Z. Zhu and G. W. Winkler contributed equally to this work.}
\author{QuanSheng Wu$^{2}$}
\author{Ju Li$^{3}$}
\author{Alexey A. Soluyanov$^{2,4}$}
\affiliation{$^{1}$Frontier Institute of Science and Technology, and State Key Laboratory for Mechanical Behavior of Materials, Xi'an Jiaotong University, Xi'an 710049, People's Republic of China}
\affiliation{$^{2}$Theoretical Physics and Station Q Zurich, ETH Zurich, 8093 Zurich, Switzerland}
\affiliation{$^{3}$Department of Nuclear Science and Engineering and Department of Materials Science and Engineering, Massachusetts Institute of Technology, Cambridge, Massachusetts 02139, USA}
\affiliation{$^{4}$Department of Physics, St. Petersburg State University, St. Petersburg, 199034 Russia}
\date{\today}
\title{Triple Point Topological Metals}

\begin{abstract}
Topologically  protected fermionic quasiparticles appear in metals, where band degeneracies occur at the Fermi level, dictated by the band structure topology. While in some metals these quasiparticles are direct analogues of elementary fermionic particles of the relativistic quantum field theory, other metals can have symmetries that give rise to quasiparticles, fundamentally different from those known in high-energy physics. Here we report on a new type of topological quasiparticles -- triple point fermions -- realized in metals with symmorphic crystal structure, which host crossings of three bands in the vicinity of the Fermi level protected by point group symmetries. We find two topologically different types of triple point fermions, both distinct from any other topological quasiparticles reported to date. We provide examples of existing materials that host triple point fermions of both types, and discuss a variety of physical phenomena associated with these quasiparticles, such as the occurrence of topological surface Fermi arcs, transport anomalies and topological Lifshitz transitions.
\end{abstract}
\maketitle

\section{Introduction}

Materials with non-trivial band structure topology, apart from possible technological applications,   provide a test ground for the concepts of fundamental physics theories in relatively cheap condensed matter experiments. For example, the recent discovery of Weyl semimetals in TaAs materials class~\cite{weng2015weyl, hasan_nat_comm, Xu_TaAs, wang_TaAs, Lv_TaAs} provided materials, where two bands cross linearly at isolated points in momentum space, called Weyl points (WPs)~\cite{Vishwanath}. These WPs occur close to the Fermi level, and hence the low energy excitations in these metals are described by the Weyl equation of the relativistic quantum field theory, thus allowing for experimental studies of Weyl fermions, examples of which in high-energy physics are still lacking.

Another example of a topological material hosting a quasiparticle analogue of an elementary particle is that of Dirac semimetals~\cite{wang_dirac_2012,wang_cd3as2,cd3as2_experiment,Xu_dirac_arpes}. These are centrosymmetric non-magnetic materials that host Dirac points (DPs) -- points of linear crossing of two doubly degenerate bands in momentum space. When DPs are located close to the Fermi level, the low energy excitations of the hosting metal are described by the Dirac equation, and thus become direct analogues of Dirac electrons in high-energy theories.

More recently, it was shown that a variety of possible symmetries realized in solids also allows for the existence of topological quasiparticle excitations, which do not have direct analogues in the Standard Model~\cite{alexey_weyl, hourglass, kim2015dirac, nodal_chains, kane_newfermions, bernevig_triple_2016, burkov_balents}, rendering novel physical behavior to the hosting compounds. Classification and description of possible topologically protected quasiparticles in solids, along with the identification of material candidates, becomes of major importance for the progress in materials science and technology, as well as in general condensed matter theory.

Several of the newly predicted topological fermionic quasiparticles appear in crystal structures that belong to non-symmorphic space groups, containing symmetries combined of a point group symmetry operation followed by translation by a fraction of the primitive unit cell vector~\cite{Vishwanath-Parameswaran, hourglass, kane_newfermions, bernevig_triple_2016, nodal_chains}. However, topological quasiparticles hosted by symmorphic space groups that contain point group symmetry operations only, are also not fully classified to date.

\begin{figure}
  \includegraphics[width = \linewidth]{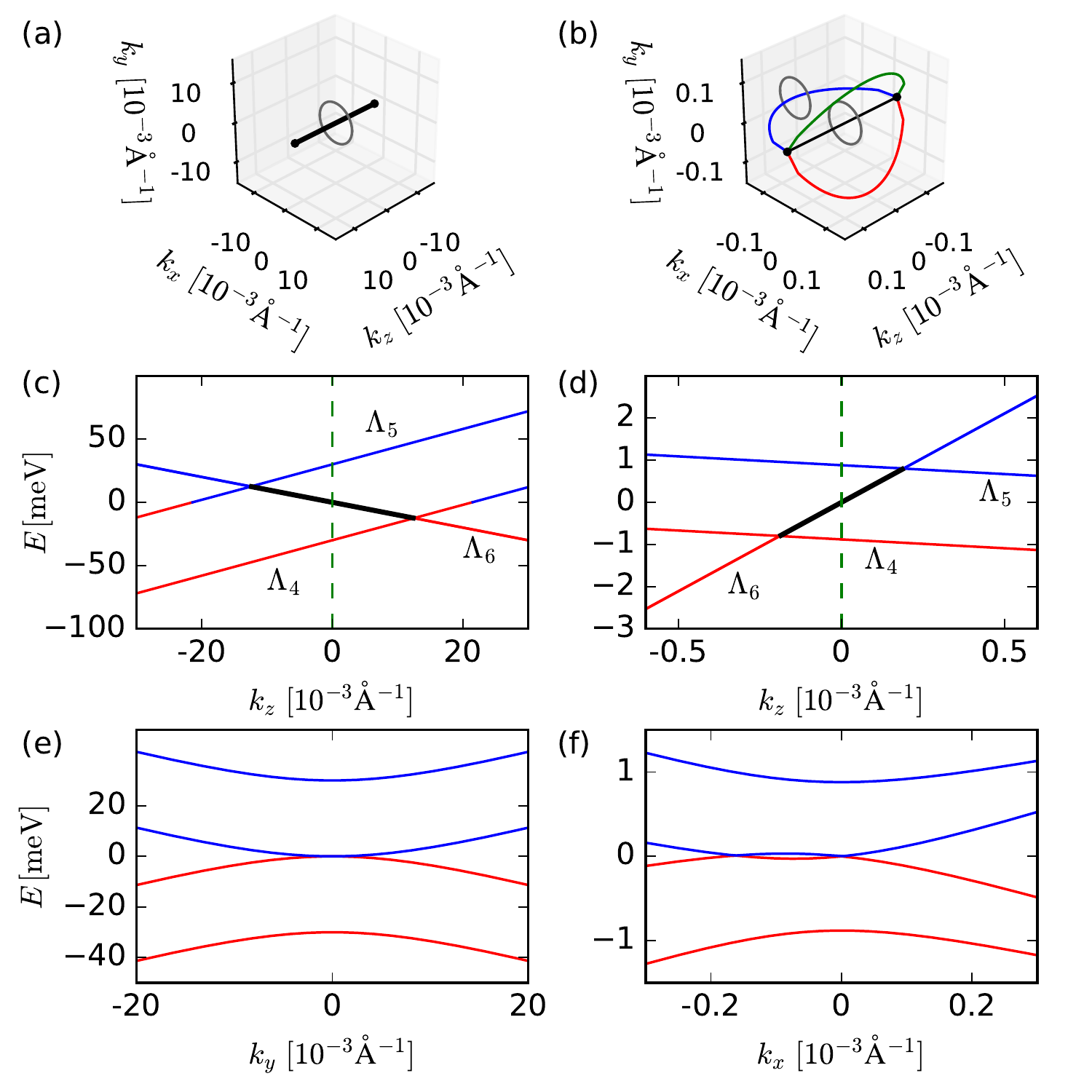}
  \caption{Two types of triple point quasiparticles. (a) Type-A triple points are connected by a single nodal line, where conduction and valence bands are degenerate (shown in black). (b) Type-B triple points are accompanied by four such nodal lines, shown in black, green, blue and red. The latter three occur in the mirror-symmetric planes in momentum space. The grey circles in (a) and (b) indicate paths for the Berry phase calculation. (c)((d)) Band structure around a type-A (type-B) triple point along the $C_3$ axis. Here $\Lambda_6$ represents the double degenerate band (double representation of $C_{3v}$), while $\Lambda_{4,5}$ correspond to two one-dimensional representations.   The black lines in (c) and (d) mark the region of the band structure that produces the nodal lines shown in black in panels (a) and (b). (e)((f)) Band structure around a type-A (type-B) triple point in a mirror symmetric plane orthogonal to $k_z$. The dashed green lines in (c) and (d) mark the momentum $k_z$ used in panels (e) and (f). Red (blue) color in panels (c-f) corresponds to occupied (unoccupied) bands.
  }
  \label{fig:nodal_lines}
\end{figure}

Here we report on a triple point (TP) fermionic quasiparticle that is realized in metallic band structures as a topologically protected crossing point of three bands, two of which are degenerate along a high-symmetry direction in momentum space. Being topologically distinct from the previously discussed three-band crossings occurring in non-symmorphic crystal structures~\cite{bernevig_triple_2016}, this TP fermion appears in {\it symmorphic} structures, the list of which is provided below. We also find several non-symmorphic space groups allowing for TP fermions, where the symmetry conditions for the appearance of TPs coincide with those of symmorphic space groups. 
TP fermions come in two topologically different variants accompanied by either one (type-A) or four (type-B) nodal lines, along which the valence and conduction bands of a metal are degenerate, as illustrated in Fig.~\ref{fig:nodal_lines}~\footnote{The TP fermions of Ref.~\onlinecite{bernevig_triple_2016}, appearing in space group 199 and 214, are topologically distinct from ours due to the fact that they are characterized by a Chern number of two and a zero dimensional Fermi surface. Ref.~\onlinecite{bernevig_triple_2016} also predicts TP fermions in space group 220, which bear some similarity to our type-B TP fermions but are fixed to a high symmetry point (see Appendix~\ref{sec:cubic}).}.

Both types of TP fermions produce topologically protected Fermi arcs on surfaces of the hosting TP topological metals (TPTMs) and have gapless Landau level spectrum when subject to symmetry-preserving magnetic fields, suggesting the possible observation of transport anomalies in these materials. Moreover, we predict series of doping-driven topological Lifshitz transitions in TPTMs, and their transition to a Weyl semimetal phase under certain lattice distortions.

We predict type-A TPTM phase to be realized in ZrTe family of compounds, where, in some cases, TPs come to interplay with other topological features of the band structure, allowing for an experimental study of coexisting topological quasiparticles. Type-B TP is realized in CuPt-ordered InAs$_{0.5}$Sb$_{0.5}$~\cite{winkler2016topological}, as well as in HgTe, strained along the (111)-direction~\cite{zaheer_hgte}. We also provide a list of space groups that can host candidate materials better suited for the experimental verification of the type-B TPTM phase. 

The paper is organized as follows. In Sec.~\ref{sec:2} we describe the conditions for the appearance of the two types of TPs, describe the topological difference between them, provide a list of hosting space groups, and describe the transition to the Weyl semimetal phase. Sec.~\ref{sec:3} provides microscopic model Hamiltonians for a generic TP, which is used to describe the topological Lifshitz transitions in TPTMs. In Sec.~\ref{sec:4} we introduce a family of experimentally known materials that host the type-A TPTM phase and provide a detailed description of their band structures. We also use the example of the predicted family of compounds to discuss the topological surface states arising in TPTMs and the response of these materials to external magnetic fields.

\section{Classification of symmorphic Triple Points}
\label{sec:2}


The realization of a symmorphic TP at a momentum $\boldsymbol{k}$ in the Brillouin zone (BZ) of a crystal structure requires the little group of $\boldsymbol{k}$ to contain both one- and two-dimensional double group representations. Thus, TPs appear on high-symmetry lines in the BZ, where the little group of $\boldsymbol{k}$ is $C_{3v}$, whose elements are 3-fold rotation $C_3$ and 3 mirrors $\sigma_v$, containing the $C_3$ axis, rotated by 120 degrees relative to each other~\cite{koster1963properties,bilbao,bradley2010mathematical}. (We also found one notable exception from this rule given by space group 174 ($C^1_{3h}$), where the interplay of time-reversal and mirror symmetry on the $C_3$-symmetric line allows for both one- and two-dimensional double group representations.) This criterion allows us to identify all the space groups that can host TP fermions on a line. 

The results are summarized in Tab.~\ref{table:summary}. Note, that the little group on the high-symmetry axis of the type-B TPTMs is exactly $C_{3v}$, while for type-A TPTMs it is supplemented by an additional anti-unitary symmetry. This symmetry is the product of the mirror plane $\sigma_h$, orthogonal to the $C_3$-axis and time-reversal (TR). Its presence preserves the existence of doubly- and singly-degenerate representations, and, hence, allows for the existence of TPs. In our consideration we also included non-symmorphic space groups, such that  the TP crossing includes the same irreducible representations as found in the symmorphic space groups.
%
\begin{table}  
  \begin{tabular}{ | c || c | c |}
    \hline
  TP type & $\Gamma$-A ($\Delta$) or $\Gamma$-P$_2$ ($\Lambda$) & K-H ($P$) or P$_0$-T ($P$)\\
  \hline
  type-A & 174, 187-190 &  \\
  \hline
  type-B & 156-161  & 157, 159-161, 183-186, 189-190 \\
  \hline
\end{tabular}
  \caption{Space groups allowing for TPs of different types with time-reversal symmetry. The points can appear on high-symmetry lines in the Brilloiun zone: $\Gamma$-A line ($\Delta=(0,0,\alpha)$) and K-H line ($P=(-1/3,2/3,\alpha)$). Commonly used notations change for the space groups 160 and 161, for which the lines (points) are $\Gamma$-P$_2$ ($\Lambda=(\alpha,\alpha,\alpha)$) and, in case the lattice constants fulfill $\sqrt{3}a>\sqrt{2}c$, also P$_0$-T ($P=(1/2-\alpha,1/2-\alpha,-1/2-\alpha)$). Note that the space groups 158, 159, 161, 184-186, 188 and 190 are non-symmorphic but the TPs exist on lines where the non-symmorphicity does not change the irreducible representations, thus they are identical to the TPs found in symmorphic space groups. We note that, in addition, the groups 162-167 and 191-194 admit type-B TPs provided time-reversal symmetry is broken in a way preserving $C_{3v}$ representations on a line in the Brillouin zone. Some of the cubic space groups also allow triple band crossings, but due to the cubic symmetry, a special situation arises which is discussed in more detail in Appendix~\ref{sec:cubic}.}
  \label{table:summary}
\end{table}

The topological classification of TPs into type-A and type-B stems from the different numbers of accompanying nodal lines, and also from the fact that the nodal lines accompanying the two types of TPs are topologically distinct. Due to the three vertical mirror planes, the Berry phase $\varphi_{\rm B}$ accumulated by valence bands on any mirror-symmetric path (shown in grey in Fig.~\ref{fig:nodal_lines}(a-b)) enclosing the corresponding nodal line is quantized to be either $0$ or $\pi$~\cite{alexandradinata_inversion_2014,schnyder_line_nodes}. The nodal line of type-A TP topological metals (TPTMs) has $\varphi_{\rm B}=0$, while all the lines of type-B TPs have $\varphi_{\rm B}=\pi$.

These values are consistent with the band structure plots, shown in Fig.~\ref{fig:nodal_lines}(c-f). In type-A TPTMs the crossing of conduction and valence (occupied and unoccupied) bands occurs on a high-symmetry line and is quadratic, while for the type-B phase this quadratic touching point splits into two points, where the bands cross linearly. The presence of nodal lines with nontrivial Berry phase, as is the case for type-B, is generally associated with the appearance of surface states~\cite{weng_drumhead,schnyder_line_nodes}. The merging and subsequent annihilation of nodal lines is similar to the nexus point discussed in the context of $^3$He-A and Bernal-stacked graphite with neglected spin-orbit coupling (SOC)~\cite{volovik_book,mikitik_sharlai,nexus1,nexus2}. We stress, however, that the scenarios discussed in the present work take full account of SOC.

Analogous to WPs~\cite{Vishwanath}, the minimal number of TPs in the BZ is four for materials preserving time-reversal symmetry. A pair of TPs located on a $C_{3v}$-symmetric line can be split into four WPs by lowering the $C_{3v}$ symmetry to $C_3$ (breaking $\sigma_v$), which can be done by a small Zeeman field parallel to the $C_3$ axis or by an atomic distortion. Conversely, imposing inversion symmetry onto the atomic structure makes the two TPs merge into a single DP. Hence, the TPTMs can be viewed as an intermediate phase separating Dirac and Weyl semimetals in materials with a $C_{3v}$-symmetric line in the BZ.

\section{Microscopic models and Lifshitz transitions}
\label{sec:3}


To analyze the physical properties of TP fermions, we now introduce $\kp$ models for TPTMs. An example of such a model for a type-B TP was provided in Ref.~\cite{winkler2016topological} for the CuPt-ordered InAs$_{0.5}$Sb$_{0.5}$ (space group 160 $C_{3v}^5$). Here we concentrate on the type-A TP, which can for example be realized in $D_{3h}$ (space groups 187 and 189) according to Tab.~\ref{table:summary}.

\begin{figure}
  \includegraphics[width = \linewidth]{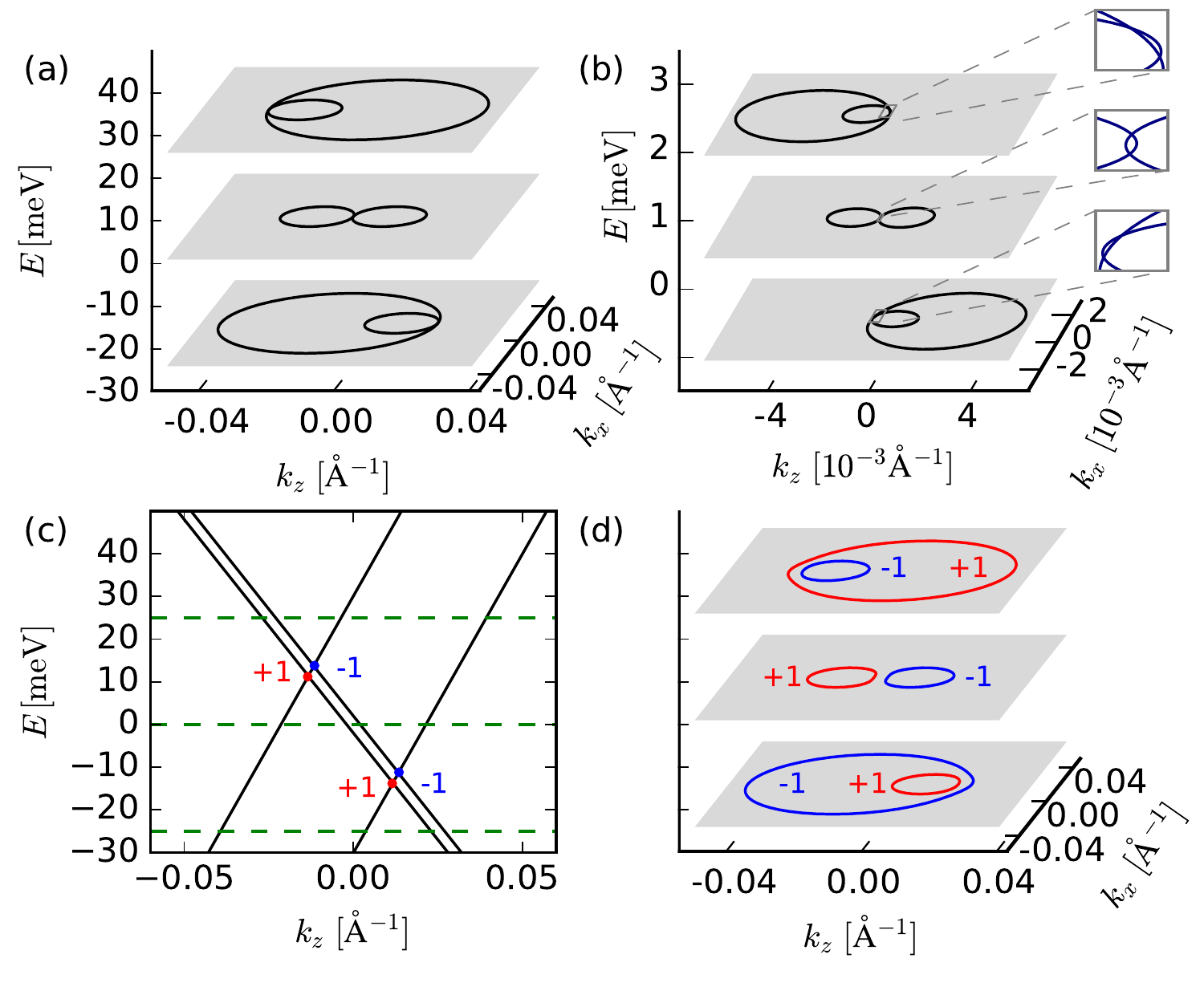}
  \caption{(a)((b)) Fermi surfaces for type-A (type-B) triple point topological metals at three different energy cuts: below, between  and above the two triple points. The three small insets in panel (b) show that for the type-B scenario there are several distinct touching points between the Fermi pockets. (c) Band structure around the triple points for a small Zeeman field parallel to the $C_3$ axis. (d) Fermi surface of type-A triple point topological metal with a small Zeeman field. In panel (c) and (d) the Chern numbers of WPs and Fermi surfaces are marked in red ($+1$) and blue ($-1$).}
  \label{fig:fermi_simple}
\end{figure}

This space groups has all the  symmetries of $C_{3v}$ supplemented by a mirror $\sigma_h$ that is orthogonal to the three-fold axis. Combined with TR $\theta$, this symmetry changes the little group of $C_{3v}$ by adding the following anti-unitary symmetry $\theta \circ \sigma_h$: $(k_x,k_y,k_z) \rightarrow (-k_x,-k_y,k_z)$, where $k_z$ is aligned with the $C_{3v}$ axis. Note, that $\theta \circ \sigma_h$ acts similar to a two-fold axis (although an anti-unitary one) and thus makes the 4 nodal lines scenario of type-B TPs incompatible with the $D_{3h}$ point group in nonmagnetic materials. As a result, the $\kp$ model for the type-A TP is different from that of type-B, and can be written as (see Appendix~\ref{sec:kp} for the model derivation)
\begin{equation}
  \scriptsize
  H_{\kp}^{\mathrm{TP}_\mathrm{A}} = \begin{pmatrix}
    E_0 + A_1 k_z  & 0 & -i\omega C k_x & i\omega C k_y\\
    0 & -E_0 +  A_2 k_z & -\omega D k_y & -\omega D k_x\\
    i\omega^* C k_x& -\omega^* D k_y & B k_z & 0 \\
    -i\omega^* C k_y& -\omega^* D k_x & 0 & B k_z \\
  \end{pmatrix},
  \label{eq:kpc}
\end{equation}
where \mbox{$\omega = -1+\sqrt{2}-i$}. In the following we used $E_0 = 30\,\mathrm{meV}$, $A_1 = A_2 = 1.4\,eV\mathrm{\AA}$, $B=-1.0\,eV\mathrm{\AA}$ and $C = D = 1.0\,eV\mathrm{\AA}$. We use above $\kp$ model for the illustrations of the type-A TPTM and the model of Ref.~\cite{winkler2016topological} for the type-B illustrations in Fig.~\ref{fig:nodal_lines} and~\ref{fig:fermi_simple}. Further details of $\kp$-modeling  of TPTMs, including the model for the type-B TP, can be found in Appendix~\ref{sec:kp}.


Using the models of Eq.~\eqref{eq:kpc} and the one of Ref.~\cite{winkler2016topological} we can analyze the Lifshitz transitions in the TPTMs. The nodal lines of Fig.~\ref{fig:nodal_lines}(a-b) guarantee that several Fermi surfaces touch within a finite energy window in between two TPs. Fig.~\ref{fig:fermi_simple}(a) illustrates the fixed $k_y=0$ cuts of the Fermi surface for the Fermi level $E_{\rm F}$ placed above, below and in between the two TPs, representing three topologically distinct Fermi surfaces. At each of the two TPs a topological Lifshitz transition takes place: one of the Fermi pockets shrinks to a point reopening either inside or outside another Fermi pocket. When the $E_{\rm F}$ is placed in between the two TPs there appears a topologically protected touching point between electron and hole pockets, similar to the type-II WP scenario~\cite{alexey_weyl}. In the Appendix~\ref{sec:fermizrte} we provide a real material illustration of the Lifshitz transitions in type-A TPTM.

The Lifshitz transitions occurring in type-B TPTMs are illustrated in Fig.~\ref{fig:fermi_simple}(b). The difference to the type-A transitions is that a single touching point between the Fermi pockets (the point of quadratic band touching) now splits into four points (or two linear band touchings on each mirror plane) due to the breaking of $\sigma_h$ (see insets of Fig.~\ref{fig:fermi_simple}(b)). Interesting spin textures with changing winding numbers, were predicted for a (111)-strained HgTe in Ref.~\cite{zaheer_hgte}, which according to our classification is a type-B TPTM. We verified that similar nontrivial windings in the spin-texture are found for type-A TPTMs.

Since the distinct Fermi pockets touch in TPTMs for a range of energies, the topological charge of individual pockets is undefined. However, as mentioned above, this degeneracy is lifted by breaking $\sigma_v$ by, for example, applying a small Zeeman field in the $z$ direction. In this case each of the TPs splits into two WPs with opposite Chern numbers as illustrated in Fig.~\ref{fig:fermi_simple}(c). The touching Fermi pockets now separate, and well-defined Chern numbers can be assigned to each of them. The Chern number of a pocket is equal to the total Chern number of WPs enclosed within it. Appendix~\ref{sec:wilson} also contains an additional topological characterization of TPs in terms of Wilson loops~\cite{yu_wilson} and Wannier charge centers~\cite{alexey_z2,gresch}.

\section{Material candidates for type-A triple point topological metal}
\label{sec:4}

Having established the physical phenomena inherent to TPSMs, we proceed to real material examples. We will use these examples to illustrate the topological surface states present in TPTMs and the non-trivial structure of Landau levels. While the material example of Type-B TPSMs was predicted to exist in CuPt-ordered InAs$_{0.5}$Sb$_{0.5}$~\cite{winkler2016topological} and (without referring to non-trivial band structure topology and topological surface states) strained HgTe~\cite{zaheer_hgte}, here we provide a list of material candidates for type-A TPTMs that were not discussed to date.

\begin{figure}[!htp]
\centerline{\includegraphics[clip, bb = 80 115 786 450,width=8cm]{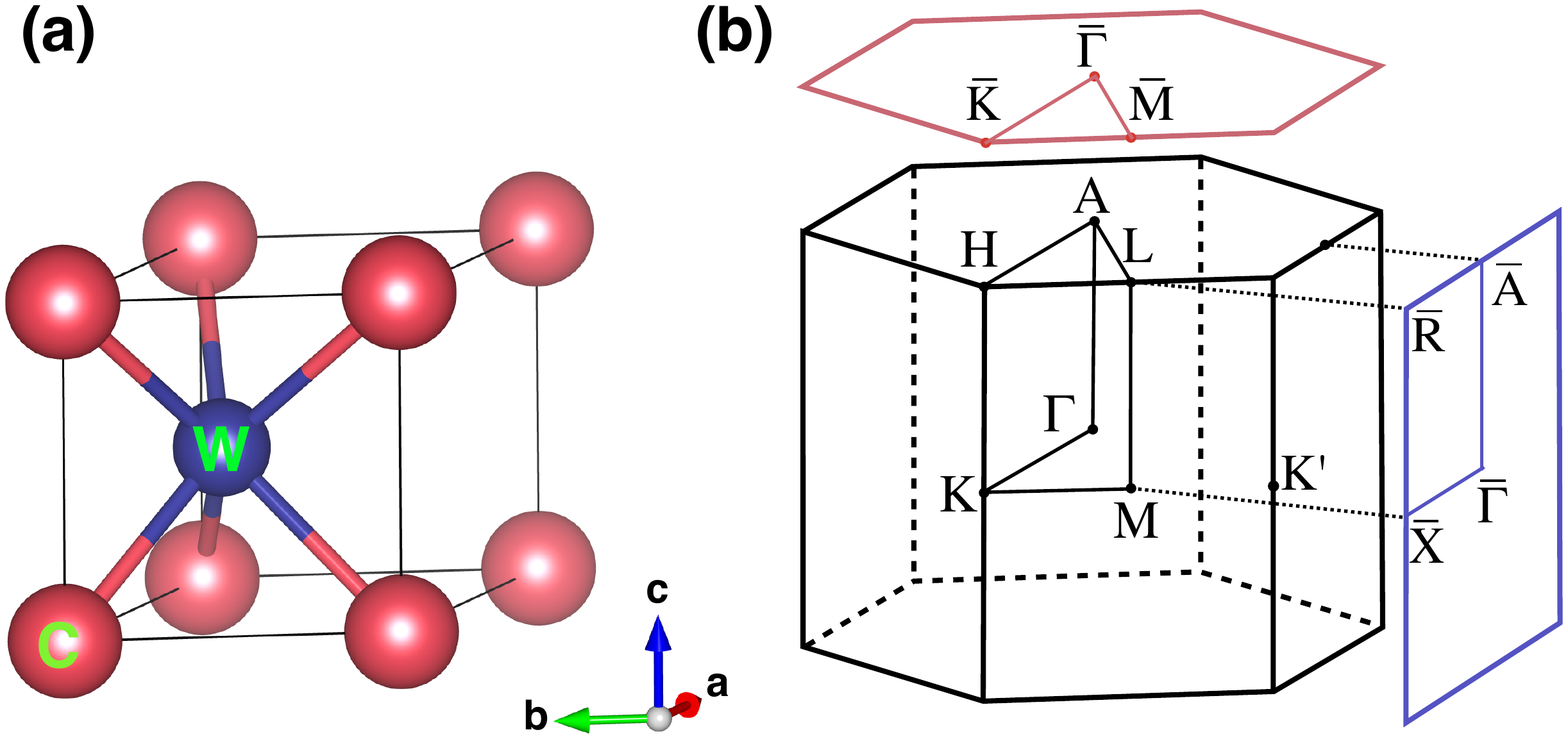}}
\caption{\label{fig:struct}
(a) Primitive unit cell of WC-type structure. (b) The bulk BZ and (001) and (010) surface BZs.}
\end{figure}

We find the type-A TPTM phase in a family of two-element metals AB (A=\{Zr, Nb, Mo, Ta, W\}, B=\{C, N, P, S, Te\}) listed in Table.~\ref{tab:materials}. These materials have a WC-type structure that belongs to space group $P\bar{6}m2$ ($D_{3h}^1$=187 ). The primitive unit cell, shown in Fig.~\ref{fig:struct}(a), consists of two atoms A and B at Wyckoff positions 1a $(0,0,0)$ and 1d $(\frac{1}{3},\frac{2}{3},\frac{1}{2})$ respectively. The corresponding bulk BZ is shown in Fig.~\ref{fig:struct}(b) along with the (001) and (010) surface BZs.

\begin{table}
\begin{tabular}{lcccc}
\hline
\hline
Material & $C_{\pm i}(k_z = 0)$ & $C_{\pm i}(k_z = \pi)$ & $E(G_1)$ [eV] & $E(G_2)$ [eV] \\
\hline
\hline
MoC\cite{Clougherty1961_564} & nodal line &  $\mp 1$ & \phantom{-}0.5119 & -0.5723 \\
WC\cite{pasquazziincorporation} & nodal line & $\mp 1$ & \phantom{-}0.3571 & -0.3286 \\
WN\cite{schonberg1954tungsten} & nodal line & $\mp 1$ & -1.2801 & \phantom{-}1.0544 \\
ZrTe\cite{orlygsson2001structure} & $\mp 1$ & $\mp 1$ & \phantom{-}0.0885 & \phantom{-}0.0438 \\
MoP\cite{boller1965kristallchemische} & $\mp 1$ & $\mp 1$ & -0.2400 & -0.3707 \\
MoN\cite{ganin2006synthesis,PhysRevB.76.134109} & $\phantom{\mp}0$ & $\mp 1$ & -1.3724 & \phantom{-}0.5984 \\
TaN\cite{schonberg1954tungsten} & $\phantom{\mp}0$ & $\mp 1$ & \phantom{-}0.0632 & \phantom{-}0.2404 \\
NbN\cite{schonberg1954tungsten} & $\phantom{\mp}0$ & $\mp 1$ & \phantom{-}0.1825 & \phantom{-}0.1513 \\
NbS\cite{Slovyanskikh1984_799} & $\mp 1$ & $\mp 1$ & -1.0090 & \phantom{-}0.2601
\end{tabular}
\caption{\label{tab:materials}A list of type-A triple point topological metal candidates. Mirror Chern numbers $C_{m=\pm i}$ in the $k_z = 0$ and $k_z=\pi$ planes are listed along with the energies of TPs relative to the Fermi level. We define $G_1$ and $G_2$ to be a pair of TPs closest to the Fermi level. References to experimental works reporting lattice parameters for the compounds are also listed.}
\end{table}

\subsection{Band structures with and without SOC}

\begin{figure*}[htp!]
\centerline{\includegraphics[width = \linewidth]{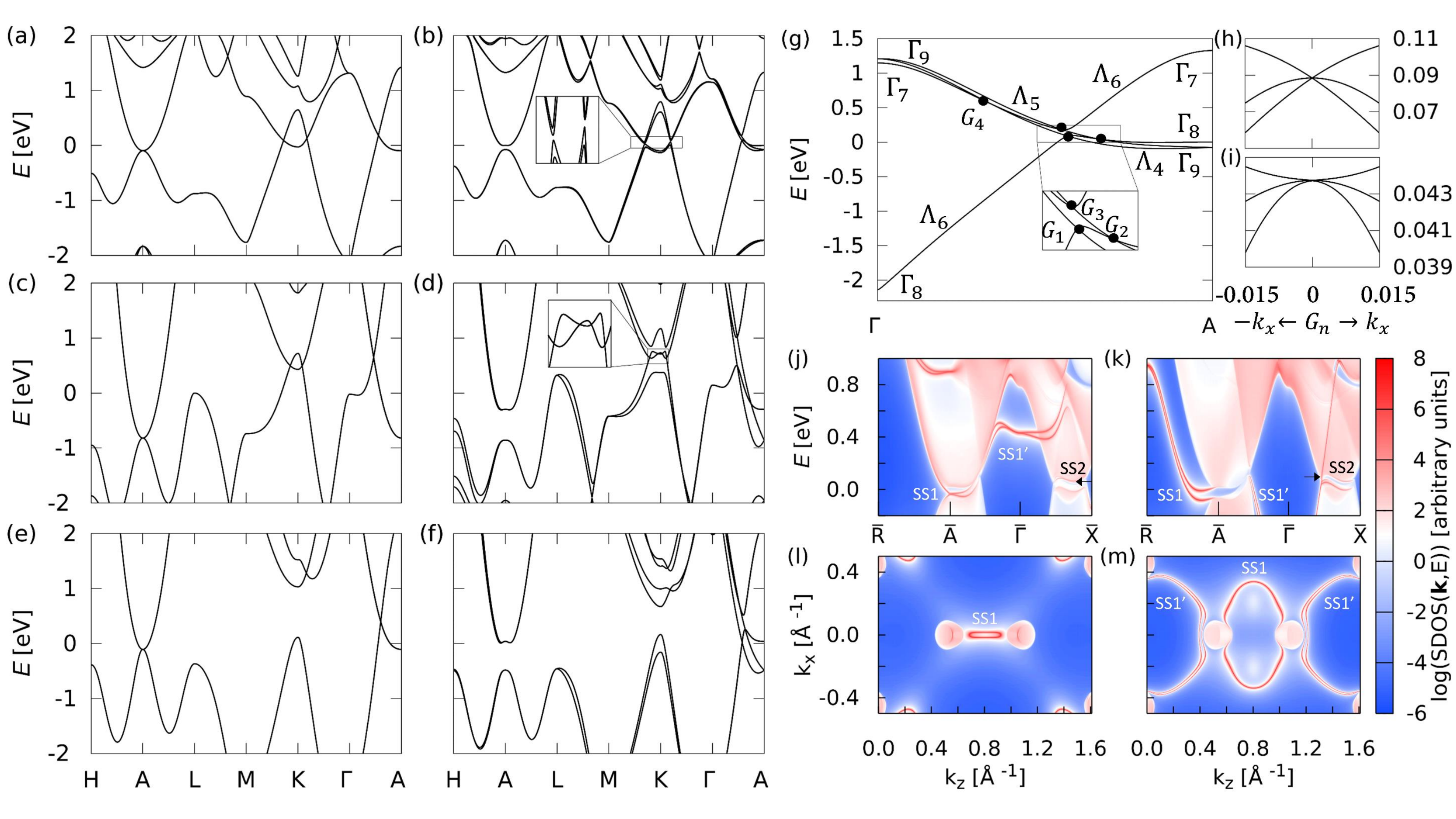}}
\caption{\label{fig:fp}
Band structure of ZrTe (a)((b)), WC (c)((d)), and TaN (e)((f)) without (with) SOC. The Fermi energy is set to 0 eV. (g) Band structure of ZrTe along the $\Gamma$-A line. Bands are labeled by their double group representations corresponding to $D_{3h}$ at $\Gamma$ and A points and $C_{3v}$ on the $\Gamma$-A line. (h)((i)) Band structures in the (100) direction with $k_z$ tuned to the TPs $G_1$ ($G_2$). (j) ((k)) Projected surface density of states (SDOS) for the (010) surface of ZrTe with Zr (Te) termination. (l)((m)) The (010)-surface Fermi surface of ZrTe at $E=0$ eV for Zr (Te) termination.}
\end{figure*}

We performed \emph{ab initio} simulations (see Appendix~\ref{sec:fp} for numerical details) of all the materials listed in Tab.~\ref{tab:materials}. For brevity, in Fig.~\ref{fig:fp} we only illustrate the band structures of ZrTe, WC and TaN, which will be used as representative materials.

In the absence of SOC there is a band inversion at K and K' points in ZrTe and WC (Fig.~\ref{fig:fp}(a) and (c)), resulting in a nodal ring in the $k_z=0$ plane protected by $\sigma_h$, while this band inversion is absent in TaN. The common feature of all materials is that along the $C_{3v}$-symmetric $\Gamma$-A line there is a band crossing of a singly and doubly degenerate bands due to the inversion of the singly- ($\Lambda_1$) and the doubly-degenerate ($\Lambda_3$) states at A. This crossing produces a single no-SOC TP, and it is this feature that generates four TPs upon introducing SOC.

All the considered materials have sizable SOC, which cannot be neglected. Due to the lack of inversion symmetry the bands are spin-split at generic momenta as shown in Fig.~\ref{fig:fp}(b,d,f). We find a band inversion along the H-A-L line such that the A point acquires an inverted gap for all materials. Consequently, the $k_z = \pi$ plane becomes an analogue of a 2D quantum spin Hall insulator in all of the compounds~\footnote{We note that we were not able to find an insulating phase in thin films of these materials}. Topological confirmation of the presence of band inversion is given by the non-trivial values of the mirror Chern numbers~\cite{mirror_chern_number} on the $\sigma_h$ plane listed in the Tab.~\ref{tab:materials} (see Appendix~\ref{sec:mirror} for more details).

In ZrTe the nodal ring around K and K' points acquires a small gap (see also the inset in Fig.~\ref{fig:fp}(b)). Interestingly, WC (together with MoC and WN) remains a nodal line metal (see inset Fig.~\ref{fig:fp}(d)). There exist two nodal rings (one inside another) formed by two touching bands protected by the horizontal mirror $\sigma_h$. For WN there is only a single such nodal ring around each $K$ and $K'$. The nodal rings are found to be quite far from the Fermi level. We find the inner (outer) nodal rings of WC 0.72~eV (0.64~eV), the inner (outer) nodal ring of MoC at 0.39~eV (0.35~eV) above the $E_{\rm F}$, while the single nodal ring of WN is 1.69 eV below $E_{\rm F}$. We specifically checked the stability of nodal lines in WC with tensile strain in the $z$ direction, and furthermore found that the inner nodal ring can be removed by applying a compressive strain of at least -0.2\% (see Appendix~\ref{sec:strain}) leaving a single nodal ring. We further calculated the band structure using the HSE06 hybrid functional~\cite{heyd2003hybrid} to check for a possible underestimation of the band gap and found that the topological features of the materials discussed above are preserved (see Appendix~\ref{sec:hybrid} for details).

In Fig.~\ref{fig:fp}(g) we show a zoom-in of the $\Gamma$-A line in ZrTe. The Fermi level resides in between the $\Gamma_9$ and $\Gamma_8$ bands at A. Upon turning on the SOC the no-SOC $\Lambda_3$ state splits into the singly degenerate $\Lambda_4+\Lambda_5$ states and the doubly degenerate $\Lambda_6$ state. Another $\Lambda_6$ state comes from the no-SOC $\Lambda_1$. The two $\Lambda_6$ states hybridize and each of them crosses with the spin-split $\Lambda_{4,5}$ states creating 2 pairs of TPs: $(G_1,G_2)$ and $(G_3,G_4)$. Each TP is protected by the $C_{3v}$ symmetry of the $\Gamma$-A line. In Fig.~\ref{fig:fp}(h-i) we show the dispersion in the (100) direction for $k_z$ tuned to the position of $G_1$, $G_2$ respectively. A linear band crossing superimposed with a quadratic band resembles a WP, degenerate with a quadratic band, similar to the findings of Ref.~\onlinecite{winkler2016topological} for type-B TPTMs.
Again, band inversion is the mechanism leading to the formation of TPs.

\subsection{Topological surface states}

The surface states  of the above compounds were calculated with the software package Wannier\_tools~\cite{wann_tools} using the symmetrized Wannier-based tight-binding model detailed in Appendix~\ref{sec:tb}, and the iterative Green's function method~\cite{sancho1984quick,sancho1985highly}. In Fig.~\ref{fig:fp}(j-k) we present the surface states of ZrTe for the (010) surface. The surface potential is found to depend strongly on the termination choice: Zr(Te)-termination is shown in Fig.~\ref{fig:fp}(j) (Fig.~\ref{fig:fp}(k)). Since the $k_z = \pi$ plane is a quantum spin Hall insulator plane, a Kramers doublet of surface states should appear along the $\overline{\mathrm{A}}$-$\overline{\mathrm{R}}$ line of the surface BZ. Indeed, we find a surface Dirac cone SS1 located at $\overline{\mathrm{A}}$ ($\overline{\mathrm{R}}$) for Zr (Te) termination. The surface states forming the Dirac cone emerge from the TPs $G_1$ and $G_2$. For $k_z$ values below the location of $G_1$ and $G_2$, there exists another pair of surface states SS1' emerging from the TPs. SS1', however, is not topologically protected, since there is no topological invariant to guarantee its appearance.

The K' point of the bulk BZ is projected onto the $\overline{\Gamma}$-$\overline{\mathrm{X}}$ line in Fig.~\ref{fig:fp}(j-k) (compare to Fig.~\ref{fig:struct}(b)). A small gap due to SOC can be visible in the projected bulk spectrum around the projection of K point (shown with an arrow).
For ZrTe the $k_z=0$ mirror plane hosts a quantum spin Hall phase with the mirror Chern numbers $\pm 1$, thus one can expect to see a Kramers pair of topological surface state along the line $\overline{\mathrm{X}}\leftarrow \overline{\Gamma}\rightarrow -\overline{\mathrm{X}}$. This expectation can be further supported by the Berry curvature calculation in the $k_z = 0$ plane. It reveals the accumulation of Berry curvature in an area around the K (K') point that sums up to approximately $-1$ ($1$) (see Appendix~\ref{sec:mirror}). In accord with this topological arguments we do find a quantum Hall like surface state SS2 crossing the gap along $\overline{\Gamma}$-$\overline{\mathrm{X}}$ (its Kramers partner is not shown, being at TR-symmetric part of the surface BZ). The choice of surface termination flips the sign of velocity of SS2. This flip is related to the fact that one of the terminations is obtained from the top surface, while the other one from the bottom one, and hence both pictures correspond to the same the surface state (the component of the $\bk$-vector orthogonal to the surface is reverted). We thus conclude that on a (010) surface there exist two topologically protected surface states, potentially observable in ARPES: SS1 and SS2.

Fig.~\ref{fig:fp}(l-m) show the (010)-surface Fermi surface revealing double Fermi arcs between the two hole pockets containing the TPs, corresponding to SS1. The state SS2 is not visible for this choice of the Fermi level. For Te termination the Fermi arcs connect the two hole pockets, while they do not touch them for the Zr termination. In both cases, however, the surface states are protected by TR and mirror symmetry of the $k_z = \pi$ plane, so they can not be fully removed from the spectrum.

While the topological protection of the Fermi arcs of Weyl semimetals in general do not rely on crystalline symmetries~\cite{Vishwanath}, the Fermi arcs of Dirac semimetals are in general not protected away from TR symmetric planes $k_z=0,\,\pi$ and can be subject to hybridization~\cite{kargarian_dirac_arc}. Still, in Dirac semimetals Cd$_3$As$_2$ and  Na$_3$Bi closed surface Fermi contours, connecting the two DPs, were found both numerically and experimentally~\cite{potter_nat_2014,wang_dirac_2012,Xu_dirac_arpes}. 
For ZrTe Fig.~\ref{fig:fp}(l) realizes an exposed closed Fermi contour scenario, showing that the Fermi arcs of TPs generally do not need to connect different non-trivial carrier pockets.

To further establish this point, we modeled the behavior of the surface states using the $\kp$ model of Appendix~\ref{sec:kp} (Eq.~\eqref{eq:kpaa} there) parametrized to have a single surface Dirac cone located at A. The $\kp$ model used allows for tuning between inversion symmetric and asymmetric  band structures (see Appendix~\ref{sec:kp}). In Fig.~\ref{fig:kp_sdos1} we compare the surface states obtained from this model with and without inversion symmetry. In the presence of inversion and TR symmetry the two TPs merge into a four-fold degenerate DP. Across all energies in the gap the two hole pockets around $\overline{\mathrm{A}}$ are connected by two Fermi arcs and the surface state on the $k_z$-axis is two-fold degenerate. Breaking of inversion symmetry then splits the DP into two TPs. Each TP contributes a single non-degenerate surface state. Since the two surface states are split along the $k_z$-axis, the Fermi arcs are not required to connect the two hole pockets. Instead, one finds a topological-insulator-like Dirac cone around $\overline{\mathrm{A}}$ which is still protected by TR and $\sigma_h$ symmetries. The splitting of the DP into two TPs thus explains the opening of Fermi arcs in ZrTe.
\begin{figure}
  \includegraphics[width = \linewidth]{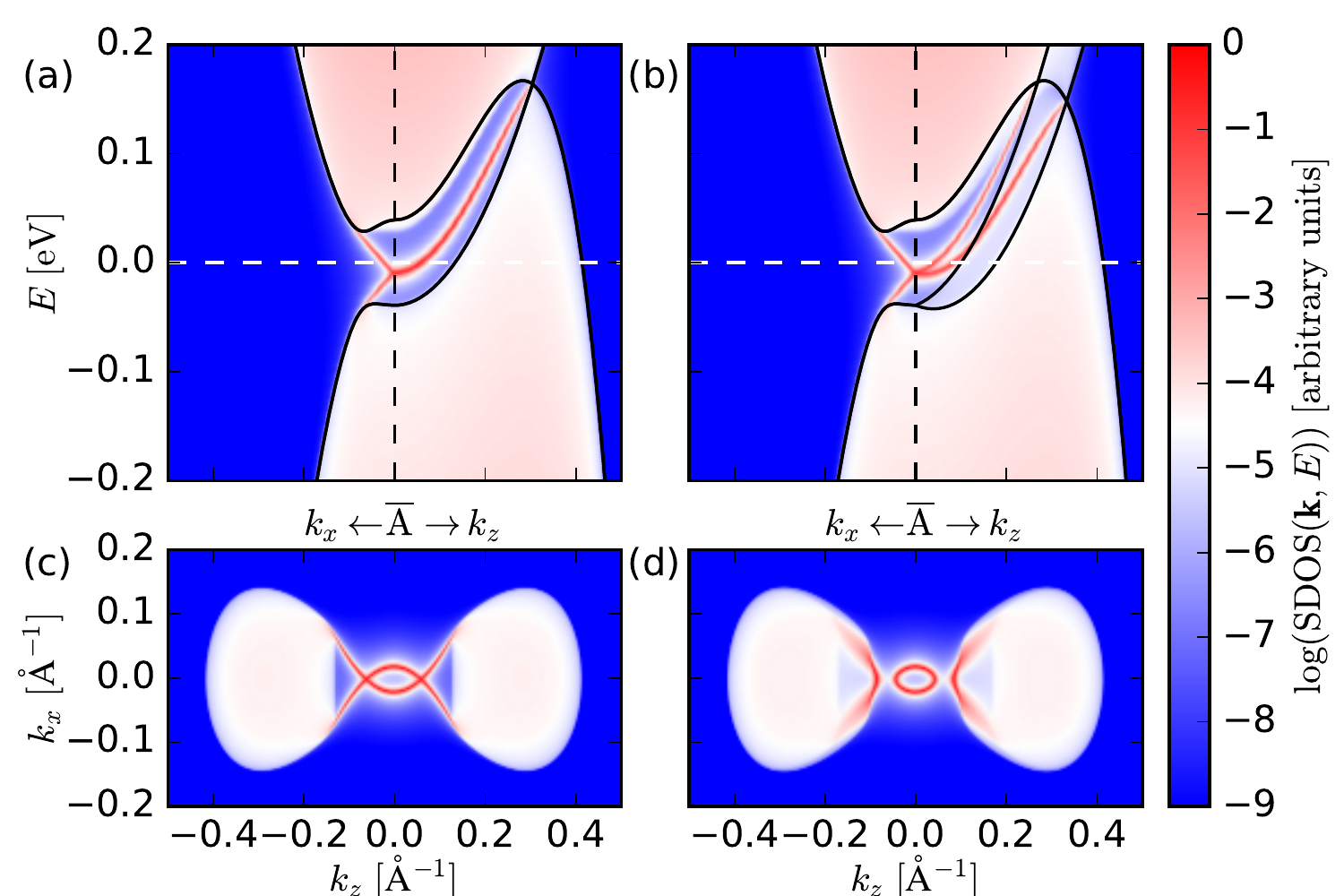}
  \caption{(010)-surface of the $\kp$ model given in Eq.~\eqref{eq:kpa}. (a) ((b)) shows the SDOS (the black lines show the bulk dispersion for $k_y=0$) and (c) ((d)) the surface Fermi surface with (without) inversion symmetry.}
  \label{fig:kp_sdos1}
\end{figure}

\subsection{Landau levels}
\label{sec:landau}
The topologically non-trivial nature of Weyl semimetals reveals itself in magnetotransport. Type-I WPs produce gapless Landau level spectrum, realizing the chiral anomaly of the quantum field theory~\cite{adler,bell-jackiw, volovik_book,nielsen_ninomiya,qi_transport,son_spivak}. Type-II WPs have an anisotropic chiral anomaly~\cite{alexey_weyl}, where the Landau level spectrum is gapless only for certain directions of the applied magnetic field.

We find that magnetotransport properties of TPs also depend on the direction of an applied magnetic field. A $C_3$ preserving magnetic field (along the $C_3$-axis) does not gap the Landau level spectrum of a TP, but instead each TP contributes a single chiral Landau level. However, if the field is applied in a $C_3$-breaking direction the Landau level spectrum becomes  gapped. Such a direction dependence also occurs in Dirac semimetals~\cite{cd3as2_landau}, further supporting the view of TPTMs as an intermediate state between Dirac and Weyl semimetals.

To illustrate our claims, we obtained the Landau levels by performing a Peierls substitution of $k_x$ and $k_y$ in the $\kp$ Hamiltonian by $k_x = \frac{i}{\sqrt{2}l_B} (a^\dag - a)$ and $k_y = \frac{i}{\sqrt{2}l_B} (a^\dag + a)$, with $l_B = \sqrt{\frac{\hbar}{eB}}$ the magnetic length and $a^\dag$, $a$ the raising and lowering operators $a^\dag |n \rangle = \sqrt{n+1} | n+1 \rangle $ and $a |n \rangle = \sqrt{n} | n-1 \rangle $. The resultant Landau level spectrum calculated with a $\kp$ model of ZrTe (see Eq.~\eqref{eq:kpa} of Appendix~\ref{sec:kp}) is shown in  Fig.~\ref{fig:landau} for a magnetic field of 20~Tesla applied in (001) direction. With the magnetic field a pair of TPs $G_1$ and $G_2$ turns into two chiral Landau levels with opposite chirality. The two chiral Landau levels are required to cross, as illustrated by the inset in Fig.~\ref{fig:landau}, resulting in gapless Landau level spectrum, that suggests strong signatures of the TPTM phase to be observable in magnetotransport.
\begin{figure}
  \includegraphics[width = \linewidth]{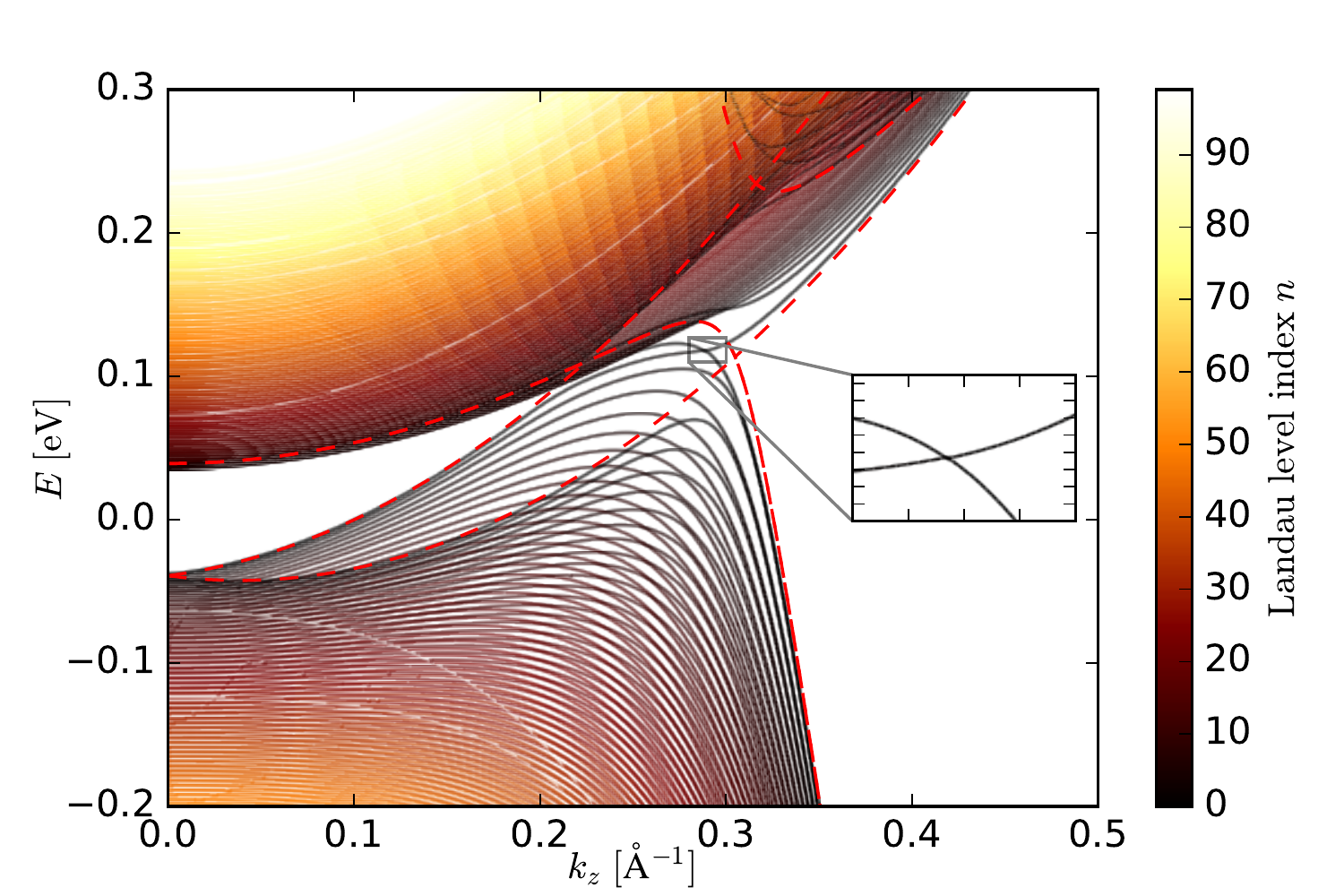}
  \caption{Landau levels for a magnetic field of 20 Tesla applied parallel to the $C_3$ axis. The  calculation is done with the $\kp$ model of Eq.~\eqref{eq:kpa} describing ZrTe. The dashed red lines show the bulk bands and the inset reveals the crossing of the two chiral Landau levels.}
  \label{fig:landau}
\end{figure}

\section{Conclusions}

In conclusion, we introduced the notion of a triple point topological metal, illustrating the topology mediated observable phenomena associated with this class of materials. We identified two topologically distinct classes of TPTMs and provided material examples for all of them. Our work also specifies the space groups that host the TPTM phase, allowing for future discovery of hosting compounds. We believe that our work will allow for further progress in understanding topological phenomena in solids, and identification of topological materials with potential applications in technology. In particular, we expect low temperature applications to arise due to the presence of direction dependent magnetotransport in TP materials.

{\it Note added:} During the completion of the manuscript, a work Ref.~\onlinecite{xi_dai_triple} appeared on arXiv, discussing some of the topological properties of TaN -- one of the materials we propose for type-A triple points in this work. 

\section{Acknowledgments}
We would like to thank T. Hyart, D. Gresch, R. Skolasinski and J. Cano for useful discussions. Z. Z. would like to specifically thank J. W. Liu for extensive discussion of the triple points in the material candidates of the paper. G. W. W., Q. S. W. and A. A. S. were supported by Microsoft Research and the Swiss National Science Foundation through the National Competence Centers in Research MARVEL and QSIT, and acknowledge M. Troyer for providing the possibility to carry out this project. Z. M. Z. and J. L. were supported by a grant-in-aid of 985 Project from Xi'an Jiaotong University. J.L. acknowledges support by NSF DMR-1410636 and DMR-1120901.
\\
\\

\appendix

\section{$\kp$ Triple points in cubic space groups}
\label{sec:cubic}
\begin{figure}
  \includegraphics[width = 0.5\linewidth]{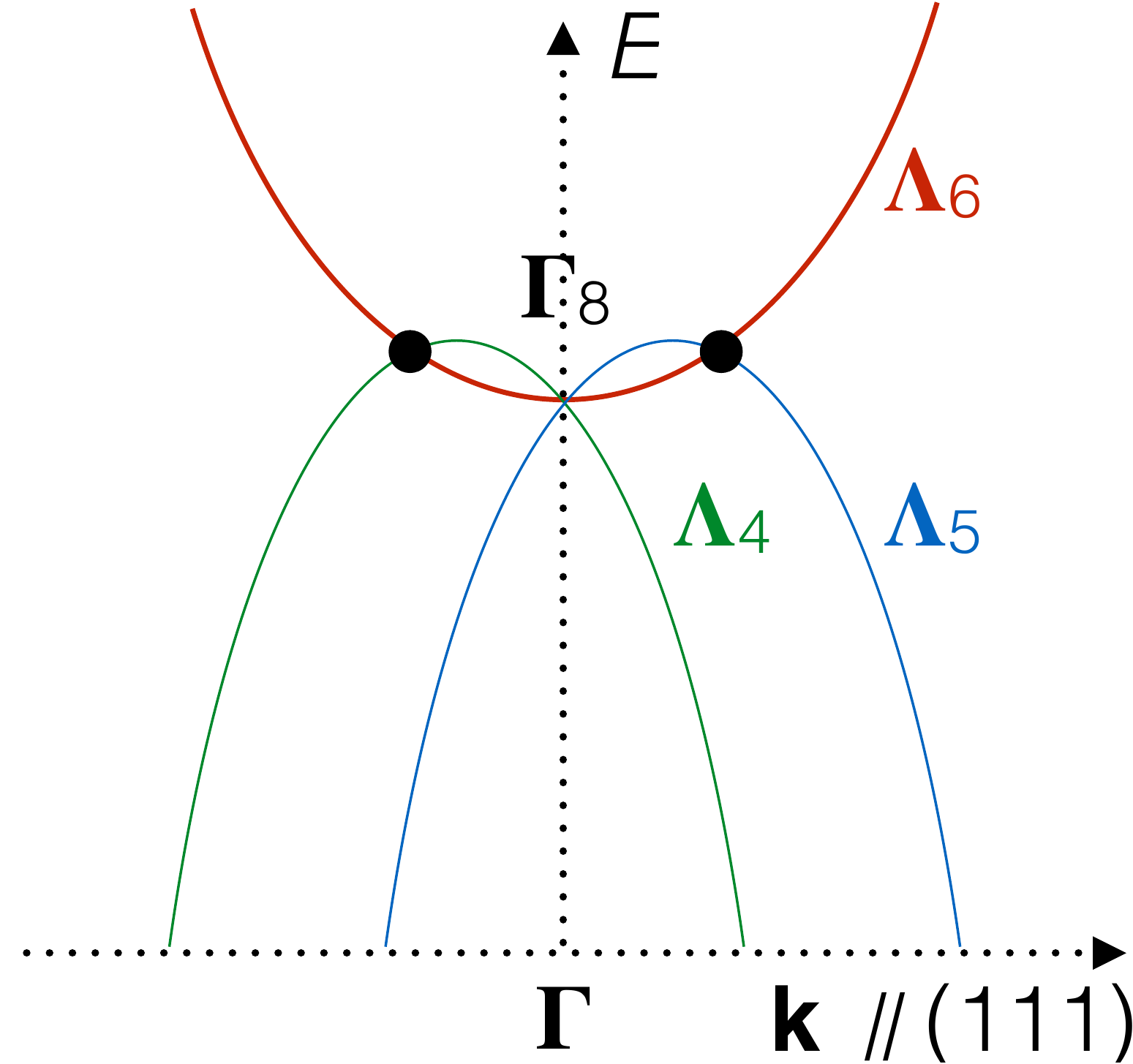}
  \caption{Schematic band structure of unstrained HgTe near $\Gamma$.}
  \label{fig:cubic}
\end{figure}
In Tab.~\ref{table:summary} we list space groups allowing for TPs of different types with time-reversal symmetry. In addition the cubic space groups 215-220 allow for type-B TPs along the $\Lambda$ high symmetry line($\Lambda=(\alpha,\alpha,\alpha)$), and for space groups 217 and 220 additionally along the $F$ line ($F=(1/4+\alpha,1/4-3\alpha,1/4+\alpha)$). The case of the non-symmorphic space group 220 has also been treated in Ref.~\onlinecite{bernevig_triple_2016}. Due to the lack of a horizontal mirror these TPs are all of type-B. The situation is complicated by the fact that the $\Lambda_{4,5}$ and $\Lambda_6$ states are degenerate at $\Gamma$ (the same is true for the H and P point connected by the $F$ line in space group 217) forming the four-dimensional $\Gamma_8$ representation. This is depicted in Fig.~\ref{fig:cubic} and leads generically to eight TPs near a $\Gamma_8$ crossing along the eight equivalent $\Lambda$ directions. It has been pointed out in Ref.~\onlinecite{zaheer_hgte} that HgTe realizes such a scenario. Since the $\Lambda_{4,5}$ and $\Lambda_6$ states are also degenerate at $\Gamma$ there is no band inversion associated with these TPs, and they correspond to the trivial scenario according to the topological classification given in Appendix~\ref{sec:wilson}. In principle the $\Lambda_{4,5}$ states could cross with a different $\Lambda_6$ state (e.g. from an electron-like $\Gamma_6$ or $\Gamma_7$ state - going into a single $\Lambda_6$ state - with lower energy than the $\Gamma_8$ state), thus generating also nontrivial TPs. We note that, in addition, the cubic space groups 221-230 admit type-B TPs provided time-reversal symmetry is broken in a way preserving $C_{3v}$ representations on a line in the Brillouin zone.

\section{$\kp$ Hamiltonians}
\label{sec:kp}
We derive several $\kp$ models describing the bands in the vicinity of the A, K and TPs. The $\kp$ models are used to get a better understanding of the surface states, Fermi surfaces and Landau levels.

\subsection{$\kp$ models around the A and triple points}
First we construct a model around the A point which captures the band inversion and describes the TPs. We include the $\Gamma_9$, $\Gamma_8$ and $\Gamma_7$ states (see Fig.~4(g) of the main text and Tab.~65 of Ref.~\onlinecite{koster1963properties}) with energies close to the Fermi level. The little group of A is $D_{3h}$ plus TR symmetry. For the derivation of the $\kp$ models we need to identify the correct representations of the symmetry operations. The Hamiltonian is then constructed such that it commutes with all symmetries $S$
\begin{equation}
  H(S(\bk)) = R_S H(\bk) R_S^\dag,
  \label{eq:symcon}
\end{equation}
with $R_S$ being the representation of S in the basis of $H$.

Since all representations are two dimensional we write the symmetry representation as the direct sum of two dimensional representations $R(\Gamma_9) \oplus R(\Gamma_7) \oplus R(\Gamma_8)$,
\begin{widetext}
\begin{equation}
\begin{aligned}
C_3 &= \begin{pmatrix} -\frac{1}{2} & -\frac{\sqrt{3}}{2} \\  \frac{\sqrt{3}}{2} & -\frac{1}{2}  \\ && 1 \end{pmatrix},\  R_{C_3} = - \mathbf{I}_{2 \times 2} \oplus \begin{pmatrix} \frac{1}{2} & \frac{-\sqrt 3}{2} \\ \frac{\sqrt 3}{2} & \frac{1}{2} \end{pmatrix} \oplus \begin{pmatrix} \frac{1}{2} & \frac{-\sqrt 3}{2} \\ \frac{\sqrt 3}{2} & \frac{1}{2} \end{pmatrix} ,\\
\sigma_v &= \mathrm{diag}\{-1,\phantom{-}1,\phantom{-}1\},\  R_{\sigma_v} = -i \tau_z \oplus \phantom{-}i \tau_z \oplus i \tau_z ,\\
\sigma_h &= \mathrm{diag}\{\phantom{-}1,\phantom{-}1,-1\},\  R_{\sigma_h} = \phantom{-}i \tau_x \oplus -i \tau_y \oplus i \tau_y ,\\
\mathrm{TR} &= \mathrm{diag}\{-1,-1,-1\},\  R_\mathrm{TR} = \begin{pmatrix} 0 & \frac{1-i}{\sqrt 2} \\ \frac{-1+i}{\sqrt 2} & 0 \end{pmatrix} \oplus -i \tau_y \oplus -i \tau_y ,
\end{aligned}
\end{equation}
\end{widetext}
with $\tau_x$, $\tau_y$ and $\tau_z$ being the Pauli matrices and $\mathbf{I}_{2\times 2}$ the identity.

Considering the constraint Eq.~\eqref{eq:symcon} for all symmetries above, one obtains the following Hamiltonian
\begin{widetext}
\begin{equation}
  {\scriptsize
  H_{\kp}^{\mathrm{A}} = \begin{pmatrix}
    \epsilon_1({\bf k}) + A k_z  & 0 & \omega E k_x & -\omega E k_y & -i\omega D k_x & i\omega D k_y\\
    0 & \epsilon_1({\bf k}) - A k_z & i \omega E k_y & i\omega E k_x & -\omega D k_y & -\omega D k_x\\
    \omega^* E k_x& -i\omega^* E k_y &  \epsilon_2({\bf k})  & 0 & -C k_y -i B k_z & C k_x \\
    -\omega^* E k_y& -i \omega^*E k_x & 0 & \epsilon_2({\bf k})  & C k_x & C k_y -i B k_z\\
    i\omega^* D k_x& -\omega^* D k_y & -C k_y + i B k_z & C k_x & \epsilon_3({\bf k}) & 0 \\
    -i\omega^* D k_y& -\omega^* D k_x & C k_x  & C k_y + i B k_z & 0 & \epsilon_3({\bf k}) \\
  \end{pmatrix},}
  \label{eq:kpa}
\end{equation}
\end{widetext}
using the definitions \mbox{$\omega = -1+\sqrt{2}-i$} and \mbox{$\epsilon_i({\bf k}) = E_i + F_i (k_x^2+k_y^2) + G_i k_z^2$} and $\bk$ relative to the $A$ point. Via fitting to the ZrTe band structure we obtain the following parameters for Eq.~\eqref{eq:kpa}: $E_1 = -0.0391$, $E_2 = 1.3709$, $E_3 = 0.0391$, $F_1 = 2.2$, $F_2 = -12.64$, $F_3 = 1.5$, $G_1=3.75$, $G_2 = -0.5$, $G_3 = 4.25$, $A=0.17$, $B=0.24$, $C=2.9$, $D=0.05$ and $E=2.55$.

Now, a minimal model, which captures the inversion of the $\Lambda_6$ and the $\Lambda_{4/5}$ states but leaves out the higher energy $\Lambda_6$ coming from the $\Gamma_9$ representation, is constructed. By removing the $\Gamma_9$ states and their interactions from Eq.~\eqref{eq:kpa} a $4\times 4$ model is obtained
\begin{equation}
  {\scriptsize
  H_{\kp}^{\mathrm{A,}4\times 4} = \begin{pmatrix}
    \epsilon_1({\bf k}) + A k_z  & 0 & -i\omega D k_x & i\omega D k_y\\
    0 & \epsilon_1({\bf k}) - A k_z & -\omega D k_y & -\omega D k_x\\
    i\omega^* D k_x& -\omega^* D k_y & \epsilon_3({\bf k}) & 0 \\
    -i\omega^* D k_y& -\omega^* D k_x & 0 & \epsilon_3({\bf k}) \\
  \end{pmatrix}.
  }
  \label{eq:kpaa}
\end{equation}
To simulate the interaction of the two $\Lambda_6$ bands we add a fourth order term to $\epsilon_3({\bf k}) = E_3 + F_3 (k_x^2+k_y^2) + G_3 k_z^2 + H_3 k_z^4$. We found the following parameters via fitting to the band structure of ZrTe: $E_1 = -0.0391$, $E_3 = 0.0391$, $F_1 = 2.2$, $F_3 = 3.2$, $G_1=4.5$, $G_3 = -7.3$,  $H_1 = 0$, $H_3 = -7.3$, $A=0.17$ and $D=0.45$. Note that the parameter $A$ is the only one that breaks inversion symmetry in the above model. Setting $A=0$ one obtains a $\kp$ description of a Dirac semimetal.

A four band Hamiltonian describing the bands in the vicinity of the two TPs can be obtained using the representations used to obtain Eq.~\eqref{eq:kpaa}. Instead of $\sigma_h$ and TR only their product $\theta \circ \sigma_h$ needs to be taken into account at a general $\bk$-point on the $C_{3v}$ axis. The resultant Hamiltonian is given in Eq.~\eqref{eq:kpc}. This is a realization of the type-A TPs introduced in the main text.

A uniform magnetic field can be added via a Zeeman term which is given in our basis as
\begin{equation}
H_\mathrm{Zeeman} = h_z \left[\tau_x \oplus (-\tau_y)\right],
\end{equation}
with $\tau_x$ and $\tau_y$ Pauli matrices and we used $h_z = 0.002$ in Fig.~3(c-d) of the main text.

The $\kp$ Hamiltonian given in Ref.~\onlinecite{winkler2016topological} is different from Eq.~\eqref{eq:kpc} due to the absence of $\sigma_h$ (or $\theta \circ \sigma_h$) symmetry in the corresponding point group. It realizes the type-B TP scenario, and is given here for completeness
\begin{equation}
  {\scriptsize
  H_{\kp}^{\mathrm{TP}_\mathrm{B}} = \begin{pmatrix}
    E_0 + A k_z & 0 & D^{\phantom{*}} k_y &  \phantom{-}D^{\phantom{*}}k_x \\
    0 & -E_0+A k_z & F^{*}k_x & -F^* k_y \\
    D^{*} k_y &  \phantom{-}F^{\phantom{*}} k_x & B k_z + C k_x & \phantom{-}C^{\phantom{*}}k_y \\
    D^* k_x & -F^{\phantom{*}}k_y & C^{\phantom{*}}k_y & B k_z-C k_x
  \end{pmatrix}.}
  \label{eq:kpc1}
\end{equation}

\subsection{$\kp$ model for the K point}
A good $\kp$ description of the topology and bands around K (or K') requires at least 8 states. The little group of the K points is $C_{3h}$ and the $\Gamma_7$, $\Gamma_{12}$, $\Gamma_{11}$, $\Gamma_9$, $\Gamma_{12}$, $\Gamma_{10}$, $\Gamma_8$ and $\Gamma_7$ states (see Tab.~57 of Ref.~\onlinecite{koster1963properties}) are determined to be relevant for constructing a $\kp$-description. We use the following symmetry representations
{
\begin{widetext}
\begin{equation}
\begin{aligned}
  C_3 &= \begin{pmatrix} -\frac{1}{2} & -\frac{\sqrt{3}}{2} \\  \frac{\sqrt{3}}{2} & -\frac{1}{2}  \\ && 1 \end{pmatrix}, R_{C_3} = \mathrm{diag}\{e^{i\frac{\pi}{3}},-1,-1,e^{i\frac{\pi}{3}},-1,e^{-i\frac{\pi}{3}},e^{-i\frac{\pi}{3}},e^{i\frac{\pi}{3}}\} ,\\
  \sigma_h &= \mathrm{diag}\{\phantom{-}1,\phantom{-}1,-1\}, R_{\sigma_h} = \mathrm{diag}\{ i,-i,i,-i,-i,i,-i,i \} .
\end{aligned}
\end{equation}
\end{widetext}
}
Considering the symmetries given above, the lowest order Hamiltonian around K is given by
\begin{widetext}
  \begin{equation}
    {\scriptsize
  H_{\kp}^\mathrm{K} = \begin{pmatrix}
  \epsilon_1({\bf k}) & 0 & B_1\,k^+ & A_1\,k_z & 0 & B_3\,k^- & 0 & 0 \\
  0 & \epsilon_2({\bf k}) & -A'_1\,k_z & B_2\,k^- & 0 & 0 & B_4\,k^+ & 0 \\
  B_1^* \, k^- & -A^{\prime *}_1\,k_z & \epsilon_3({\bf k}) & 0 & A_2\,k_z & B_9\,k^+ & 0 & B_5\,k^- \\
  A_1^*\,k_z & B_2^*\,k^+ & 0 & \epsilon_4({\bf k}) & -B'_9\,k^+ & 0 & B_6\,k^- & A_4\,k_z \\
  0 & 0 & A_2^* \,k_z & -B^{\prime *}_9\,k^- & \epsilon_5({\bf k}) & 0 & B_7\,k^+ & 0 \\
  B_3^*\,k^+ & 0 & B_9^*\,k^+ & 0 & 0 & \epsilon_6({\bf k}) & A_3\,k_z & B_8\,k^+ \\
  0 & B_4^*\,k^- & 0 & B_6^*\,k^+ & B_7^*\,k^- & A_3^*\,k_z & \epsilon_7({\bf k}) & 0 \\
  0 & 0 & B_5^*\,k^+ & A_4^*\,k_z & 0 & B_8^*\,k^- & 0 & \epsilon_8({\bf k})
  \end{pmatrix},}
  \label{eq:kpk}
\end{equation}
\end{widetext}
using $\epsilon_i$ defined as in Eq.~\eqref{eq:kpa}, $k^\pm = k_x \pm i k_y$ and $\bk$ relative to $K$. Since K is not a time-reversal invariant momentum bands do not form doubly degenerate Kramers pairs at this point. 
For the $\kp$ model around the K point  we obtain the following parameters via fitting to the ZrTe band structure: $E_1 = -0.0979$, $E_2 = -0.0671$, $E_3 = 0.6538$, $E_4 = 0.8393$, $E_5 = 1.0661$, $E_6 = 1.1351$, $E_7 = 1.2145$, $E_8 = 1.2774$, $F_1 = F_2 = 3.6$, $F_3=F_4 = -2.0$, $F_5=F_6= 6.0$, $F_7=F_8=1.5$, $G_1 = G_2 = 3.6$, $G_3 = G_4 = -0.2$, $G_5 = G_6=2.0$, $G_7=G_8=-3.0$, $A_1 = A'_1= 4.0$, $A_2 = 0.2$, $A_3 = 0$, $A_4 = 0$, $B_1 = 0.2-i 0.1$, $B_2 = 0.02-i 0.01$, $B_3 = 0.2$, $B_4 = -0.2$, $B_5 = -1.0 + i 4.0$, $B_6 = -4.0+i 1.0$, $B_7 = 3.0 + i 0.5$, $B_8 = -0.5 + i 3.0$ and $B_9 = 1.5$.

The Weyl points reported in Ref.~\onlinecite{xi_dai_ZrTe} are also described by this $\kp$ model.

\subsection{Surface states from $\kp$ models}
Here we compare surface states obtained from the $\kp$ models to the first principles results presented in Sec.~\ref{sec:4}. The $\kp$ surface state calculations are done by discretizing the momentum $k_y$, and thus generating a 1D tight-binding model with auxilliary parameters $k_x$ and $k_z$~\cite{kp_discretisation}. The SDOS is then calculated using the iterative Green's function method~\cite{sancho1984quick,sancho1985highly}. For the $\kp$ models given in Eq.~\eqref{eq:kpa} and \eqref{eq:kpaa} we use 1~\AA{} as the discretization length and 2~\AA{} for Eq.~\eqref{eq:kpk}.
\begin{figure}
  \includegraphics[width = \linewidth]{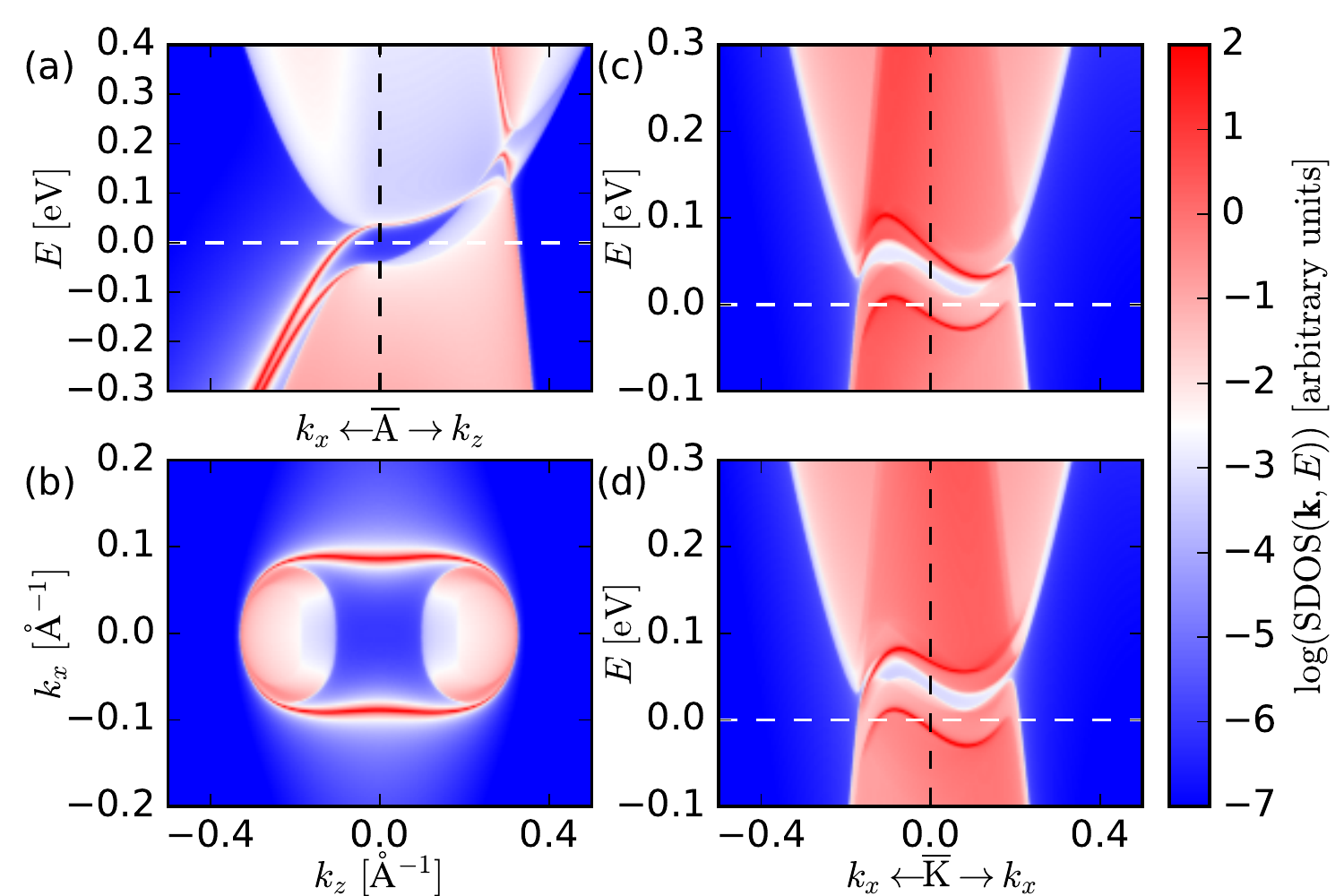}
  \caption{SDOS of the (010)-surface for the $\kp$ models given in Eq.~\eqref{eq:kpa} and \eqref{eq:kpk}. (a) and (b) show the SDOS and Fermi surface around the $\overline{\mathrm{A}}$ point. (c) ((d)) show the SDOS around the $\overline{\mathrm{K}}$ point for the top (bottom) surface.}
  \label{fig:kp_sdos}
\end{figure}

The $\kp$ models with the parameters given above fit the band structure of ZrTe. The model around A given in Eq.~\eqref{eq:kpa} is then characterized by the mirror Chern numbers $C_{m=\pm i} = \mp 1$ in the $k_z=\pi$ plane. Therefore, a topological-insulator-like surface state is expected on a surface orthogonal to the $\sigma_h$ mirror plane. In Fig.~\ref{fig:kp_sdos}(a) we show the SDOS on a surface orthogonal to $y$, corresponding to the (010) surface in the WC structure. On the $k_x$ axis the upper topologically nontrivial surface state emerges from the conduction bands and connects to the valence bands. There is another trivial surface state with opposite mirror eigenvalue below. If we compare this to the first-principles surface states shown in Fig.~\ref{fig:fp}(k) of the main text then these two surface states will form a Dirac cone at $\overline{\mathrm{R}}$ for Te-terminated surface. In Fig.~\ref{fig:kp_sdos}(b) the Fermi surface is plotted. The topologically nontrivial hole pockets are connected by a pair of Fermi arcs.

The $\kp$ model around K is characterized by a total Chern number of $C=1$, respectively $C=-1$ at K'.
Hence around K and K' a quantum Hall like surface state is expected. This is confirmed in Fig.~\ref{fig:kp_sdos}(c) and (d), where we calculated the SDOS on a surface orthogonal to $y$. The surface states give an excellent match to the first principles result presented in Fig.~4(j-k) of the main text.

\section{First-principles calculations}
\label{sec:fp}
The first-principles calculations were performed using the Vienna \emph{ab initio} simulation package (VASP)\cite{kresse1996efficient} with the projector augmented wave method~\cite{paw}. For the exchange correlation energy we considered both the generalized gradient approximation (GGA)~\cite{perdew1996generalized} within the Perdew-Burke-Ernzerhof (PBE) functional and hybrid functionals (HSE06)~\cite{heyd2003hybrid}. The energy cutoff was set to $560$ eV and a  $11\times11\times11$ Monkhorst-Pack mesh was used for the Brillouin zone integration. For the convergence of the electronic self-consistent calculations the total energy difference criterion was set to $10^{-8}$ eV. The lattice constants were fully relaxed until  the total energy is converged to $10^{-7}\,\mathrm{eV}$ and the residual forces on atoms are below $10^{-3}\,\mathrm{eV/\AA}$.

\subsection{Lattice constants}
\label{sec:lc}
In Tab.~\ref{tab:lc} we give the relaxed lattice constants $a_\mathrm{sim.}$ and $c_\mathrm{sim.}$ and compare them to experimental data $a$, $c$ where available. The lattice constants were fully relaxed until  the total energy is converged to $10^{-7}\,\mathrm{eV}$ and the residual forces on atoms are below $10^{-3}\,\mathrm{eV/\AA}$. In our work we adopted the relaxed lattice parameters $a_\mathrm{sim.}$ and $c_\mathrm{sim.}$ for all our simulations.

\begin{table}
\centering
\caption{Experimental and simulated lattice constants $a$ and $c$.}
\label{tab:lc}
\begin{tabular}{lllll} 
\hline
\hline
Materials & \begin{tabular}[c]{@{}c@{}} $a_\mathrm{exp.} \ [{\mathrm\AA}]$ \end{tabular}  & \begin{tabular}[c]{@{}c@{}}  $c_\mathrm{exp.} \ [{\mathrm\AA}]$   \end{tabular} & \begin{tabular}[c]{@{}c@{}} $a_\mathrm{sim.} \ [{\mathrm\AA}]$ \end{tabular}  & \begin{tabular}[c]{@{}c@{}}  $c_\mathrm{sim.} \ [{\mathrm\AA}]$   \end{tabular}\\
\hline
\hline
MoC\cite{Clougherty1961_564} & \begin{tabular}[c]{@{}c@{}} ${2.898} $ \end{tabular}  & \begin{tabular}[c]{@{}c@{}}  ${2.809} $   \end{tabular} & \begin{tabular}[c]{@{}c@{}} ${2.922} $ \end{tabular}  & \begin{tabular}[c]{@{}c@{}}  ${2.824} $   \end{tabular}\\
WC\cite{pasquazziincorporation} & \begin{tabular}[c]{@{}c@{}} ${2.928} $ \end{tabular} &  \begin{tabular}[c]{@{}c@{}}  ${2.835} $   \end{tabular} & \begin{tabular}[c]{@{}c@{}} ${2.906} $ \end{tabular} &  \begin{tabular}[c]{@{}c@{}}  ${2.837} $   \end{tabular}\\
WN\cite{schonberg1954tungsten} & \begin{tabular}[c]{@{}c@{}} ${2.890} $ \end{tabular}  & \begin{tabular}[c]{@{}c@{}}  ${2.830} $   \end{tabular} &\begin{tabular}[c]{@{}c@{}} ${2.873} $ \end{tabular}  & \begin{tabular}[c]{@{}c@{}}  ${2.922} $   \end{tabular}\\
ZrTe\cite{orlygsson2001structure}  & \begin{tabular}[c]{@{}c@{}} ${3.771} $ \end{tabular}  & \begin{tabular}[c]{@{}c@{}}  ${3.861} $   \end{tabular} & \begin{tabular}[c]{@{}c@{}} ${3.800} $ \end{tabular}  & \begin{tabular}[c]{@{}c@{}}  ${3.903} $   \end{tabular}\\
MoP\cite{boller1965kristallchemische} & \begin{tabular}[c]{@{}c@{}} ${3.220} $ \end{tabular}  & \begin{tabular}[c]{@{}c@{}}  ${3.190} $   \end{tabular} & \begin{tabular}[c]{@{}c@{}} ${3.256} $ \end{tabular}  & \begin{tabular}[c]{@{}c@{}}  ${3.195} $   \end{tabular}\\
MoN\cite{ganin2006synthesis,PhysRevB.76.134109}  & \begin{tabular}[c]{@{}c@{}} ${2.868} $ \end{tabular}  & \begin{tabular}[c]{@{}c@{}}  ${2.810} $   \end{tabular} & \begin{tabular}[c]{@{}c@{}} ${2.886} $ \end{tabular}  & \begin{tabular}[c]{@{}c@{}}  ${2.856} $   \end{tabular}\\
TaN\cite{schonberg1954tungsten} & \begin{tabular}[c]{@{}c@{}} ${2.930} $ \end{tabular}  & \begin{tabular}[c]{@{}c@{}}  ${2.880}  $   \end{tabular} & \begin{tabular}[c]{@{}c@{}} ${2.816} $ \end{tabular}  & \begin{tabular}[c]{@{}c@{}}  ${2.791} $   \end{tabular} \\
NbN\cite{schonberg1954tungsten} & \begin{tabular}[c]{@{}c@{}} ${2.940} $ \end{tabular}  & \begin{tabular}[c]{@{}c@{}}  ${2.790} $   \end{tabular} & \begin{tabular}[c]{@{}c@{}} ${2.976} $ \end{tabular}  & \begin{tabular}[c]{@{}c@{}}  ${2.901} $   \end{tabular} \\
NbS\cite{Slovyanskikh1984_799}  &\begin{tabular}[c]{@{}c@{}} ${3.350} $ \end{tabular}  & \begin{tabular}[c]{@{}c@{}}  ${3.200} $   \end{tabular} & \begin{tabular}[c]{@{}c@{}} ${3.267} $ \end{tabular}  & \begin{tabular}[c]{@{}c@{}}  ${3.322} $   \end{tabular} \\
\end{tabular}
\end{table}

\subsection{HSE06 band structures}
\label{sec:hybrid}
\begin{figure}
 \includegraphics[width = 0.8\linewidth]{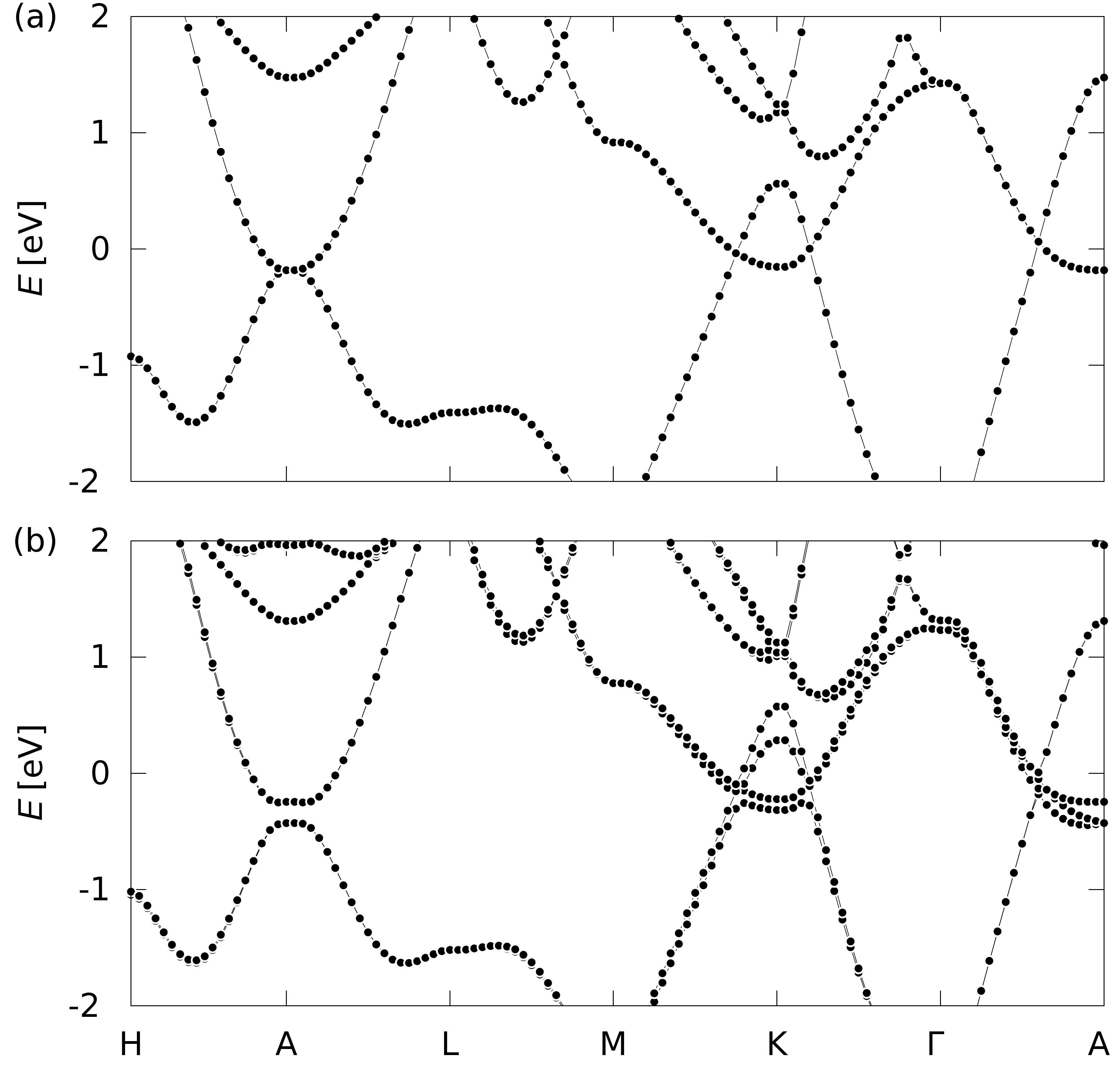}
 \caption{(a)((b)) Band structures of ZrTe without (with) SOC using hybrid functional.}
 \label{fig:hse_bs}
\end{figure}

In Fig.~\ref{fig:hse_bs} we show the band structure of ZrTe calculated using the implementation of the HSE06 hybrid functional in VASP~\cite{heyd2003hybrid,heyd_hybrid_2004,heyd_hybrid_2006}. Compared to Fig.~4(a-b) of the main text, we find all important features of the GGA calculation, i.e. band inversion at A and K and the existence of four TPs.

\subsection{Stability of nodal rings in WC}
\label{sec:strain}
\begin{figure}
 \includegraphics[width = 0.8\linewidth]{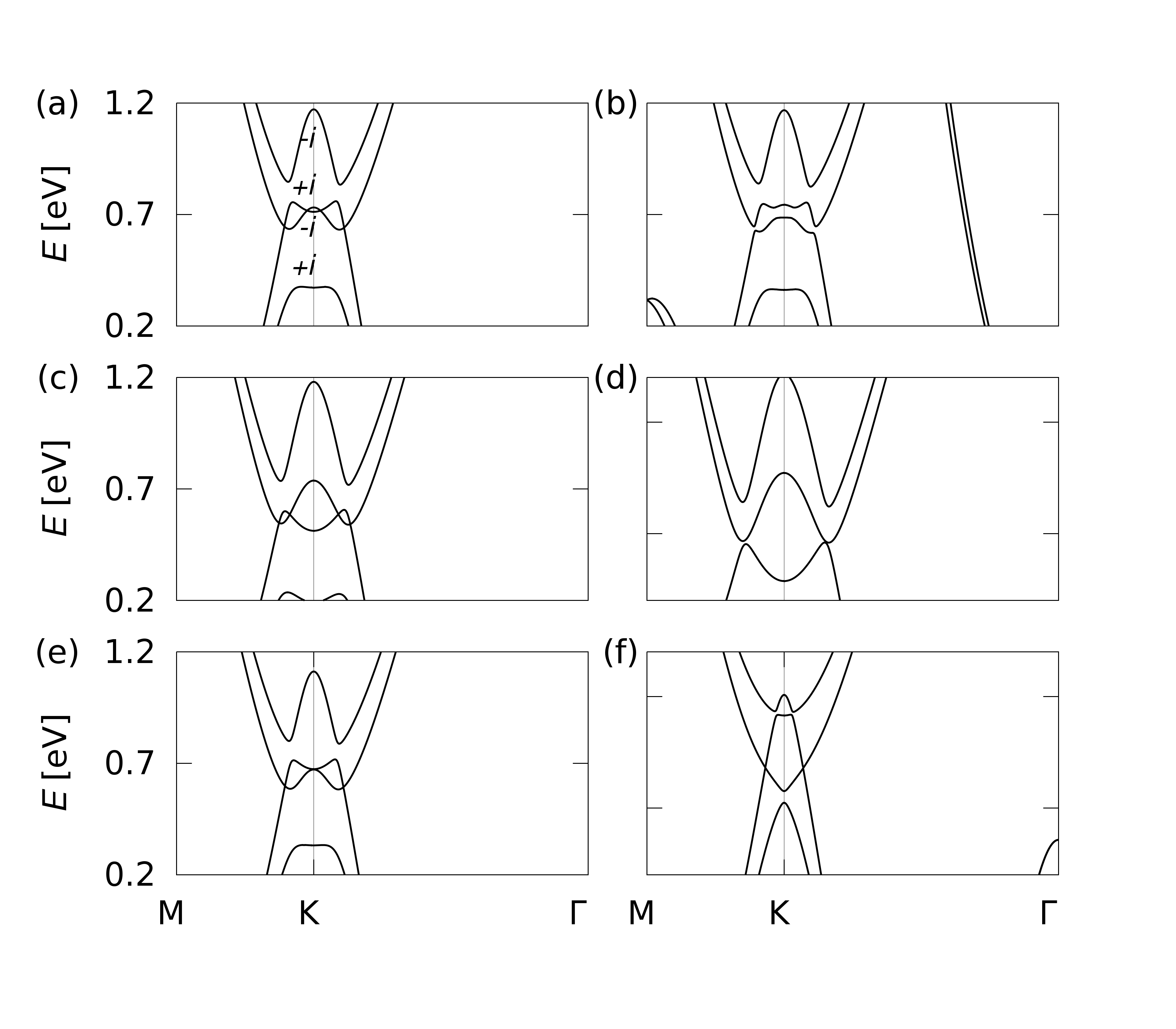}
 \caption{(a)((b)) Band structure of WC with (without) horizontal mirror symmetry. In the case with mirror symmetry the mirror eigenvalues of the bands are given. (c)((d)) Band structure of WC with tensile strain of $2\%$ ($5\%$), (e)((f)) with compressive strain of $-0.2\%$ ($-3\%$) along z direction.}
 \label{fig:transition}
 \end{figure}

In Fig.~\ref{fig:transition}(a) we show a zoom-in of the band structure in the $k_z=0$ plane around the K point. Bands are marked by their horizontal mirror $\sigma_v$ eigenvalue $\pm i$. Breaking the horizontal mirror, by moving the C atom by $0.01 \,\mathrm{\AA}$ in the $z$ direction, gaps the nodal rings (Fig.~\ref{fig:transition}(b)).

In Fig.~\ref{fig:transition}(c-f) we test the stability of the nodal rings under (001) strain. We find that the double nodal ring survives up to $2\%$ tensile strain. $-0.2\%$ compressive strain leaves a single nodal ring which survives up to about $-3\%$ compressive strain.

\section{Effective Hamiltonian from Wannier projection}
\label{sec:tb}
The surface state calculation and topological classifications are usually illustrated with effective tight-binding (TB) Hamiltonians generated from the first-principles Wannier functions~\cite{mostofi2008wannier90,weng2009revisiting}. For the materials discussed in this work we projected the first-principle wavefunctions on \emph{s}, \emph{p} and \emph{d} orbitals located at site A and \emph{p} orbitals at site B, without performing the iterative spread minimization. For ZrTe we choose the lower (upper) bound of the outer energy window for the disentanglement as $0.0$ eV ($21.0$ eV), and the bottom (top) of frozen energy window as $0.0$ eV ($12.3$ eV). The obtained atomic-like Wannier functions were used to construct then a 24-band (including spin) TB Hamiltonian, which reproduces the first-principles band structures with sub-meV accuracy.

One major issue of Wannier derived TB Hamiltonians is that the Wannier functions do not exactly fulfill all crystal symmetries. One consequence is that symmetry protected band crossings will therefore always appear as avoided crossings with a sub-meV gap. Several works deal with this problem during the wannierization process~\cite{sakuma,selectively_localized_wannier,alexey_wire}, but we found that a post-processing approach gives very good results in our case. We imposed the three point group and TR symmetries via calculating the group average of the TB Hamiltonian
\begin{equation}
H(\bk) = \frac{1}{|G|} \sum_{g \in G} R_g H(g(\bk)) R_g^{-1}
\end{equation}
with $G$ the symmetry group containing $|G|$ elements $g$, and $R_g$ the representation of $g$ for atomic wavefunctions. The prerequisite for this approach to work is, of course, that the Wannier functions transform similar to atomic wavefunctions. We used this symmetrized TB for calculating surface states, mirror Chern numbers and the Wilson loop characterization of the TPs introduced in Sec.~\ref{sec:wilson}.

\section{Fermi surface of ZrTe}
\label{sec:fermizrte}
\begin{figure}
  \includegraphics[width = 0.8\linewidth]{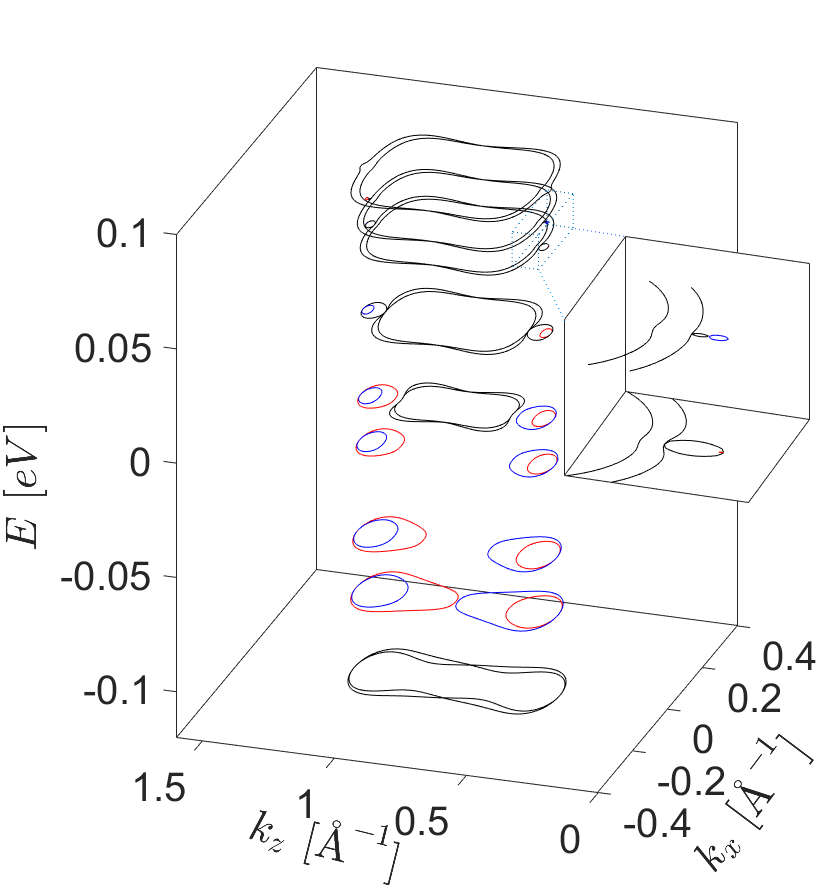}
  \caption{Fermi surface of ZrTe at different energies. We assign a topological charge to each surface defined for an infinitesimal magnetic field in the $z$ direction. Red (blue) corresponds to Chern number +1 (-1) and black is Chern number 0. The inset is a zoom-in of the second and third highest energies shown.}
  \label{fig:fermi}
\end{figure}

In the main text we discussed the generic Fermi surface and Lifshitz transitions connected to TPs. We find that ZrTe is an excellent platform for studying our predictions in a real material.

In Fig.~\ref{fig:fermi} we show the Fermi surface of ZrTe at different energies. At the Fermi level there are four Fermi surfaces centered around A (neglecting possible Fermi surfaces around K and K'). We assign topological charges according to the scenario that an infinitesimal magnetic Zeeman field is applied in the $z$ direction. Upon lowering the energy from the Fermi level, topologically nontrivial hole pockets touch and their topological charges annihilate. Raising the energy, one approaches the TPs $G_1$ and $G_2$. At the lower energy TP $G_2$ the outer hole pocket touches with the electron pockets centered around A. The electron pockets connect the two nontrivial hole pockets opposite of A and their topological charges annihilate. At the higher energy TP $G_1$ the inner hole pocket reduces to a point and then reappears outside as an electron pocket with opposite topological charge (see inset of Fig.~\ref{fig:fermi}). Increasing the energy further only the electron pockets centered at A remain.

\section{Mirror Chern numbers}
\label{sec:mirror}
\begin{figure}[htp]
  \includegraphics[width = \linewidth]{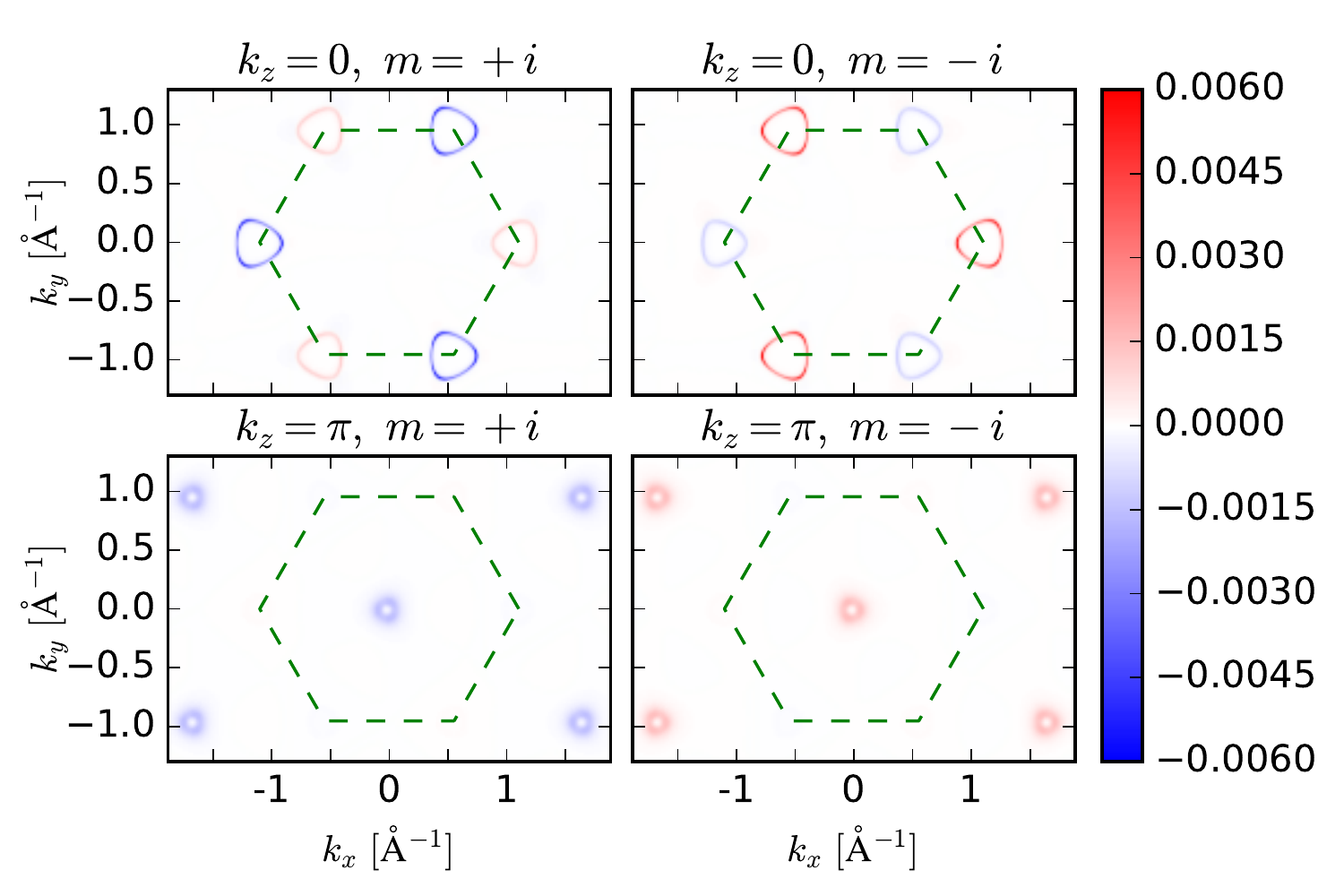}
  \caption{The Berry curvature for specific mirror eigenvalues on $\sigma_h$-mirror invariant planes in ZrTe.}
  \label{fig:mirror}
\end{figure}

The nontrivial topology of ZrTe is driven by band inversions at the A, K and K' points. These points are located in the $k_z =\pi$ and $k_z = 0$ planes, which are both invariant under the horizontal $\sigma_h$ mirror operation. This enables us to plot the Berry curvature for specific mirror eigenvalues $m=\pm i$ on these planes as shown in Fig.~\ref{fig:mirror}~\cite{mirror_chern_number,snte_pred}. We facilitate the mirror Chern number calculations with the symmetrized tight-binding models.

In ZrTe both planes $k_z = 0,\, \pi$ are characterized by mirror Chern numbers $C_{m=\pm i} = \mp 1$ (see Tab.~I in the main text). Fig.~\ref{fig:mirror} clearly shows that the areas of high Berry curvature are localized around the A, K and K' points, which confirms the band inversion at these points. Due to the mirror Chern numbers we expect TI-like surface states on any surface perpendicular to the mirror plane. These surface states are protected by mirror and TR symmetry. If the crossing points on the $\Gamma$-A line are opened by sufficient $C_3$-symmetry breaking strain the bulk becomes insulating and the above mirror Chern numbers lead to a weak topological insulator phase. The materials TaN, MoN, and NbN, with trivial mirror Chern numbers in the $k_z = 0$ plane, become strong topological insulators for sufficient $C_3$-symmetry breaking strain.

\section{Wilson loop characterization for pairs of triple points}
\label{sec:wilson}
The Wilson loop can be defined on any path in $k$-space connecting two points ${\bf k_1}$ and ${\bf k_2}$ with the property ${\bf k_1} = {\bf k_2 } + {\bf G}$, where ${\bf G}$ is a reciprocal lattice vector. The Wilson loop is defined as the path ordered product~\cite{yu_wilson}
\begin{equation}
\mathcal{W}_{\bf k_1 k_2} = P_{\bf k_1} \left( \prod_{j=1,2,\dots} P_{\bf k'_j} \right) P_{\bf k_2},
\label{eq:wilson}
\end{equation}
with \mbox{$P_{\bf k} = \sum_{n\in \mathrm{occ.}} |u_n({\bf k})\rangle \langle u_n({\bf k}) |$} the projector on the occupied subspace of a Hamiltonian. The Wilson loop is inherently gauge invariant due to the gauge invariance of the projector $P_{\bf k}$. The Berry phase associated with the loop is given by the determinant of the Wilson operator \mbox{$\det(\mathcal{W}) = \exp(i\phi_B)$}. If the Hamiltonian has a symmetry $R$, it can be shown that~\cite{andrei_point_group}
\begin{equation}
\tilde R \mathcal{W}_{\bf k_1 k_2} \tilde R^{-1} = \mathcal{W}_{R{\bf k_1}R{\bf k_2}},
\label{eq:wsym}
\end{equation}
with $R$ acting in reciprocal space and $\tilde R$ in occupied band space. The symmetry expectation value of Wilson loop eigenstates $|v_i \rangle$ is calculated as $\langle v_i | \tilde R | v_i \rangle$.

We show here that a pair of TPs may be characterized by a $\mathbb Z_2$ topological invariant. For Weyl~\cite{alexey_weyl} and Dirac~\cite{gresch} semimetals it is known that the Wilson loop spectrum on a sphere enclosing the semimetallic point gives the topological classification of the crossing. Also in our case with TPs a similar kind of topological classification is possible. We apply the classification to the symmetrized tight-binding model for ZrTe (see Appendix~\ref{sec:tb}).
\begin{figure}
  \includegraphics[width =  \linewidth]{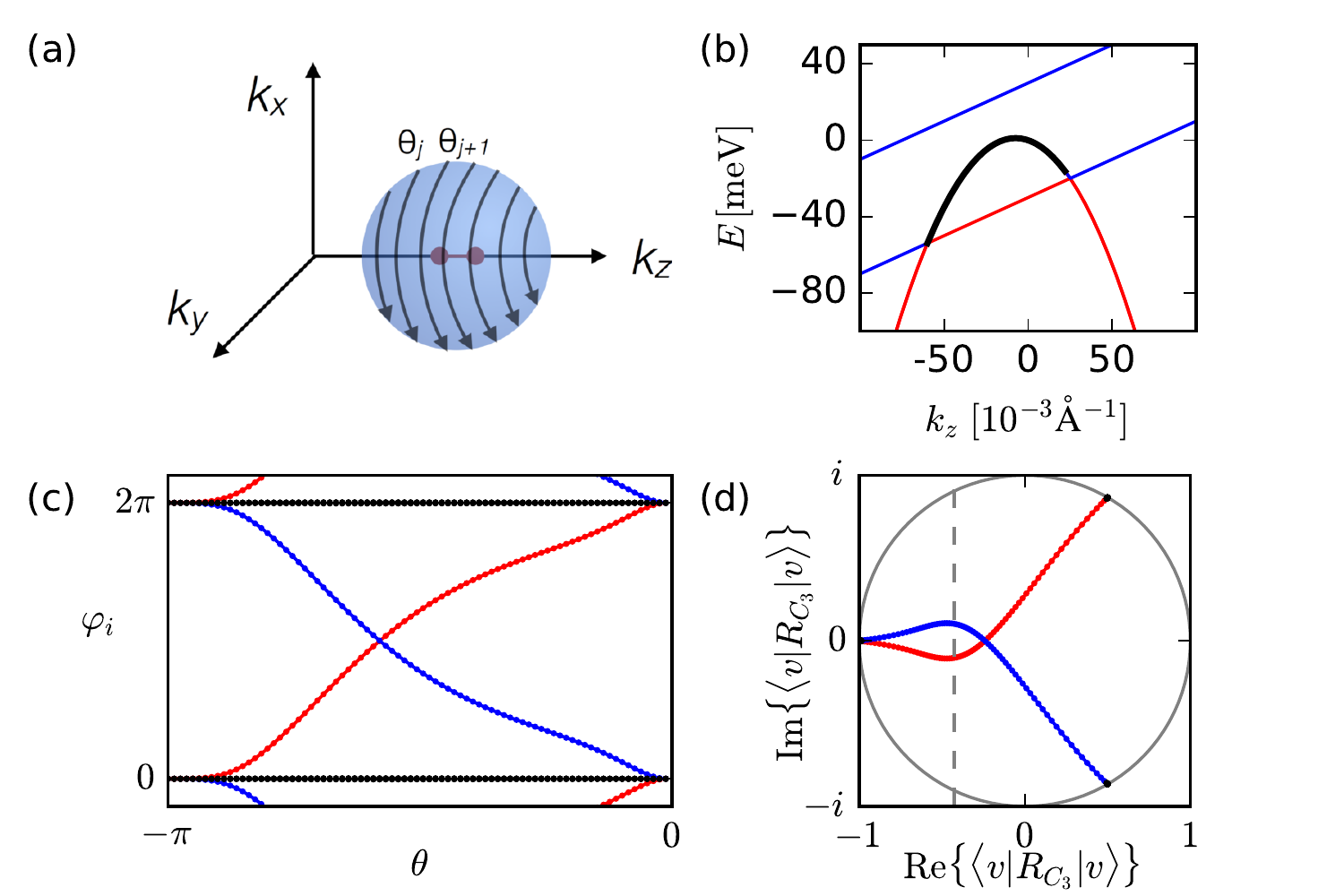}
  \caption{(a) Pair of TPs (red points) connected by a nodal line (red line) enclosed by a sphere. The arrows indicate the individual Wilson loops winding around the sphere. (b) Example of a trivial pair of TPs. (c) The Wilson loop spectrum on a sphere enclosing a pair of TPs. Two Wilson loop eigenvalues feature gapless flow (colored in red and blue). (d) $C_3$ symmetry expectation value of individual Wilson lines. }
  \label{fig:wilson}
\end{figure}

In Fig.~\ref{fig:wilson}(a) we show a spherical surface on which the Wilson loop spectrum is to be evaluated. The sphere is chosen such that the Hamiltonian is gapped everywhere on the surface, the symmetry axis containing the TPs goes through the center of the sphere and both TPs are enclosed by the sphere. The latter point is important, since there is always at least one nodal line connecting two TPs, therefore including only one TP would not fulfill the requirement that the Hamiltonian is gapped on the sphere. Note that the Wilson lines are oriented such that the symmetry axis goes through their center. In Fig.~\ref{fig:wilson}(c) we plot the phases $\phi_i$ of the individual Wilson loop eigenvalues as a function of the azimuthal angle $\theta$. The tight-binding model has 8 occupied states, therefore, we obtain 8 Wilson loop eigenvalue phases $\phi_i$. 6 $\phi_i$ (marked in black) are trivial and stay very close to 0 ($2\pi$), but two (marked in red and blue) seem to cross. Note that the $\sigma_v$ symmetry constrains the $\phi_i$ such that the Wilson loop spectrum is mirror symmetric $\phi_i = -\phi_j$~\cite{alexandradinata_inversion_2014}. Since the Hamiltonian is gapped on the surface, and the Wilson loop is gauge invariant, the individual $\phi_i$ change smoothly with $\theta$. Therefore, the connectivity of the $\phi_i$ can be determined as long as they are not degenerate. To obtain the connectivity across the degeneracy point between the red and blue Wilson eigenvalues we calculate the $C_3$ symmetry expectation values of the corresponding states in Fig.~\ref{fig:wilson}(d). The grey dashed line in Fig.~\ref{fig:wilson}(d) indicates the position of the crossing of the blue and red line in Fig.~\ref{fig:wilson}(c). Note that the crossing of red and blue lines in Fig.~\ref{fig:wilson}(d) is accidental and we found that it can be avoided via choosing a cigar-shape, rather than a sphere. However, the $C_3$ symmetry expectation value is nondegenerate at the crossing Fig.~\ref{fig:wilson}(c) and we can use Fig.~\ref{fig:wilson}(d) to unambiguously determine the connectivity for all $\theta$. Therefore, the red and blue lines in the Wilson loop spectrum clearly indicate two hidden Berry curvature fluxes, one inward and one outward, through the sphere. The fluxes can be separated in the Wilson loop eigenbasis, corresponding to individual Chern numbers~\cite{alexey_smooth} of $\pm 1$. The difference of the two individual Chern numbers divided by two constitutes a $\mathbb Z_2$ topological invariant for TPs.

At the polar regions $\theta \approx 0$ or $\theta \approx -\pi$ the Wilson loop commutes with the $C_3$ symmetry due to Eq.~\eqref{eq:wsym}. In this case the $C_3$ expectation value in Fig.~\ref{fig:wilson}(d) is one of the possible $C_3$ eigenvalues $\{-1,\exp(i\pi/3),\exp(-i\pi/3)\}$, which are the starting and ending points of the lines in Fig.~\ref{fig:wilson}(d). Note that the 6 trivial $\phi_i$ (black dots) are almost fixed to the $C_3$ eigenvalues, whereas the two nontrivial $\phi_i$ (red/blue dots) change the $C_3$ eigenvalue from $\{-1,\, -1\}$ to $\{\exp(i\pi/3),\, exp(-i\pi/3)\}$. Responsible for this behaviour are the two valence bands having the rotational eigenvalues $-1,\, -1$ for $k_z$ to the left of the two TPs and $\exp(i\pi/3),\, exp(-i\pi/3)$ for $k_z$ to the right of the two TPs. Therefore, the planes above $G_1,\, G_2$ are topologically distinct from the planes below, consequently uncovering the existence of crossing points realized as the two TPs here.

In Fig.~\ref{fig:wilson}(b) we give an example of a topologically trivial pair of TPs. In this case the $C_3$ eigenvalues of the valence bands are the same to the left and to the right of the two TPs and hence the Wilson loop spectrum is in general gapped with an even $\mathbb Z_2$ invariant.

\bibliography{refs}

\end{document}